\documentclass[usenatbib]{mnras}

\usepackage{amsmath}
\usepackage{graphicx}
\usepackage{txfonts}
\usepackage{array}
\usepackage{longtable}
\usepackage[usenames]{color}

\bibpunct{(}{)}{;}{a}{}{,}

\DeclareMathAlphabet{\mymath}{U}{eus}{m}{n}

\def\eg{{\it e.g.\ }}

\def\ie{{\it i.e.\ }}

\newcommand{\msun}{\mbox{$M_{\odot}$} }

\def\la{\mathrel{\hbox{\rlap{\hbox{\lower4pt\hbox{$\sim$}}}\hbox{$<$}}}}
\def\ga{\mathrel{\hbox{\rlap{\hbox{\lower4pt\hbox{$\sim$}}}\hbox{$>$}}}}

\newcommand{\be}{\begin{equation}}
\newcommand{\ee}{\end{equation}}

\newcommand{\bi}{\begin{itemize}}
\newcommand{\ei}{\end{itemize}}

\newcommand{\ben}{\begin{enumerate}}
\newcommand{\een}{\end{enumerate}}

\newcommand{\bfig}{\begin{figure}\begin{minipage}{140mm}}
\newcommand{\efig}{\end{minipage}\end{figure}}

\newcommand{\btab}{\begin{table}\begin{minipage}{140mm}}
\newcommand{\etab}{\end{minipage}\end{table}}

\newcommand{\bfigMore}{\begin{figure}\begin{minipage}{160mm}}
\newcommand{\efigMore}{\end{minipage}\end{figure}}

\newcommand{\btabMore}{\begin{table}\begin{minipage}{160mm}}
\newcommand{\etabMore}{\end{minipage}\end{table}}

\newcommand{\bea}{\begin{eqnarray}}
\newcommand{\eea}{\end{eqnarray}}

\newcommand{\bega}{\begin{gather}}
\newcommand{\eega}{\end{gather}}

\newcommand{\bc}{\begin{center}}
\newcommand{\ec}{\end{center}}

\newcommand{\dif}{{\rm d}}

\def\araa{ARA\&A}%
\def\apj{ApJ}%
\def\apjl{ApJ}%
\def\apjs{ApJS}%
\def\aap{A\&A}%
\def\mnras{MNRAS}%
\def\pasj{PASJ}%
\def\physrep{Phys.~Rep.}%

\newcommand*{\ltsim}{\ {\raise-.75ex\hbox{$\buildrel<\over\sim$}}\ }

\def\pz{$P\left ( z\right )$\ }

\def\fgas{$f_{\mathrm{gas}}$\ }
\def\chandra{\emph{Chandra}}
\def\rmeas{$r_{2500}$}
\def\ratiomeas{$0.956\pm 0.082$}

\title [Chandra Hydrostatic Mass Calibration] {Cosmology and astrophysics from relaxed galaxy clusters - IV: Robustly calibrating hydrostatic masses with weak lensing}

\author[D. E. Applegate et al.]
{\parbox[t]{\textwidth}{
D. E. Applegate$^{1}$
\thanks{E-mail:dapple@astro.uni-bonn.de}, 
A. Mantz$^{2,3,4,5}$,
S. W. Allen$^{4,5,6}$,
A. von der Linden$^{4,5,7}$,
R. G. Morris$^{4,6}$,
S. Hilbert$^{8,9}$,
P. L. Kelly$^{10}$,
D. L. Burke$^{4,5}$,
H. Ebeling$^{11}$,
D. A. Rapetti$^{7}$,
R. W. Schmidt$^{12}$}\\
      \vspace*{3pt}
\\
$^{1}$Argelander-Institut f{\"u}r Astronomie, Auf dem H{\"u}gel 71, D-53121 Bonn, Germany\\
$^{2}$Kavli Institute for Cosmological Physics,
University of Chicago,
5640 South Ellis Avenue,
Chicago, IL 60637-1433, USA\\
$^{3}$Department of Astronomy and Astrophysics, 
University of Chicago,
5640 South Ellis Avenue,
Chicago, IL 60637-1433, USA\\
$^{4}$Kavli Institute for Particle Astrophysics and Cosmology,
Stanford University,
452 Lomita Mall,
Stanford, CA  94305-4085, USA\\
$^{5}$Department of Physics,
Stanford University,
382 Via Pueblo Mall, 
Stanford, CA  94305-4060, USA\\
$^{6}$SLAC National Accelerator Laboratory, 
2575 Sand Hill Road, 
Menlo Park, CA 94025, USA\\
$^{7}$Dark Cosmology Centre, Niels Bohr Institute, University of Copenhagen Juliane Maries Vej 30, 2100 Copenhagen \O, Denmark\\
$^{8}$Exzellenzcluster Universe, Boltzmannstr. 2, 85748 Garching, Germany\\
$^{9}$Ludwig-Maximilians-Universit{\"a}t, Universit{\"a}ts-Sternwarte, Scheinerstr. 1, 81679 M{\"u}nchen, Germany\\
$^{10}$Department of Astronomy, University of California, B-20 Hearst Field Annex \# 3411, Berkeley, CA 94720-3411, USA\\
$^{11}$Institute for Astronomy, 
2680 Woodlawn Drive, 
Honolulu, HI 96822, USA\\
$^{12}$Astronomisches Rechen-Institut, Zentrum f{\"u}r Astronomie der Universit{\"a}t Heidelberg, M{\"o}nchhofstrasse 12-14, D-69120 Heidelberg, Germany\\
}

\begin{document}

\date{}

\pagerange{\pageref{firstpage}--\pageref{lastpage}} \pubyear{2015}

\maketitle

\label{firstpage}

\begin{abstract}
  This is the fourth in a series of papers studying the astrophysics and cosmology of massive, dynamically relaxed galaxy clusters. 
Here, we use measurements of weak gravitational lensing from the Weighing the Giants project to calibrate \chandra{} X-ray measurements of total mass that rely on the assumption of hydrostatic equilibrium. 
This comparison of X-ray and lensing masses provides a measurement of the combined bias of X-ray hydrostatic masses due to both astrophysical and instrumental sources.
  Assuming a fixed cosmology, and within a characteristic radius ($r_{2500}$) determined from the X-ray data, we measure a lensing to X-ray mass ratio of $0.96 \pm 9\% \mathrm{(stat)} \pm 9\% \mathrm{(sys)}$.
  We find no significant trends of this ratio with mass, redshift or the morphological indicators used to select the sample.
  In accordance with predictions from hydro simulations for the most massive, relaxed clusters, our results disfavor strong, tens-of-percent departures from hydrostatic equilibrium at these radii.
  In addition, we find a mean concentration of the sample measured from lensing data of $c_{200} = 3.0_{-1.8}^{+4.4}$.
  Anticipated short-term improvements in lensing systematics, and a modest expansion of the relaxed lensing sample, can easily increase the measurement precision by 30--50\%, leading to similar improvements in cosmological constraints that employ X-ray hydrostatic mass estimates, such as on $\Omega_m$ from the cluster gas mass fraction.
\end{abstract}


\section{Introduction}

Total masses of galaxy clusters are of central importance in multiple cosmological measurements, notably those involving the cluster baryon fraction and mass function (for a review, see \eg \citealt{aem11}). 
Historically, much of this work has used  mass estimates from X-ray data, ultimately relying on the assumption of hydrostatic equilibrium. 
In general, for a randomly selected cluster and arbitrary measurement radius, simulations predict that departures from equilibrium will induce bias at the tens of per cent level between such estimates and the true cluster masses \citep{nagai07, lau09}. 
Restricting the analysis to intermediate measurement radii and the most massive, dynamically relaxed clusters reduces the expected bias and scatter considerably ($\ltsim 10$ per cent). 
The small intrinsic scatter seen in gas mass fractions, \fgas, for such clusters (e.g., Paper~II of this series, \citealt{mantz_fgas}) verifies that hydrostatic X-ray mass estimates trace the true mass well at these radii for these clusters. 
However, some overall average bias may remain, whether due to residual non-thermal support or instrument calibration \mbox{\citep{ndg10, schellenberger14, nelson14}}. 
Determining this bias requires independent mass measurements for the same clusters using an unbiased method.

Weak gravitational lensing (WL) provides a direct probe of the gravitational potential of a cluster, independent of baryonic physics. 
While WL mass estimates are inherently noisy due to cluster triaxiality and line-of-sight structure,  they can in principle provide unbiased mass estimates on average \citep{bas01, saasfee}.
In practice,  systematic effects must be carefully controlled, as demonstrated by the Weighing the Giants project \citep[WtG;][]{paper1,paper2,paper3}. 
But, with this in hand, WL can provide the necessary overall calibration of X-ray hydrostatic mass estimates.
A growing body of work has used WL to either directly calibrate cosmological tests or to understand the biases in published studies \citep{anja_planck, israel14, paper4, hoekstra2015}. 
In addition to enhancing cluster cosmology directly, measurements of the lensing to X-ray mass ratio constrain the net effect of astrophysical and instrumental biases on the X-ray measurements\footnote{Unless otherwise explicitly stated, the term ``X-ray masses'' refer to hydrostatic mass estimates.}.

Several groups have previously compared lensing masses to X-ray masses measured in various ways \citep[\eg][]{vbe09, zhang10, planck_i3, mahdavi13, israel14, donahue14}. 
However, those efforts have used different X-ray telescopes, different calibration versions for those telescopes, varying sample selection, and varying lensing and X-ray analysis methods. 
As a result, it is difficult to draw conclusions any broader than the relative calibration of each individual set of WL and X-ray mass estimates. 
This only underscores the need for cluster mass estimation with strict tolerances on systematic uncertainties, which WL can, in principle, provide.

This is the fourth in a series of papers dedicated to studying the cosmology and astrophysics enabled by X-ray and robust lensing observations of dynamically relaxed clusters.
Paper~I \citep{mantz_morph} details the morphological selection of dynamically relaxed clusters from a sizable subset of cluster observations in the \chandra{} data archive\footnote{\url{http://cxc.harvard.edu/cda}} and the X-ray data processing  of the sample.
Paper~II \citep{mantz_fgas} derives cosmological constraints from the sample, using measurements of gas mass fractions, while Paper~III \citep{mantz_profiles} studies the thermodynamics and scaling relations of these clusters. 
The current paper uses the 12 clusters in common between our relaxed sample and WtG to constrain the ratio of lensing to X-ray mass estimates, a calibration that was used for the cosmological measurements in Paper~II, and that impacts directly on the constraints on the cosmic mean matter density, $\Omega_\mathrm{m}$.

\label{sec:contents}
The paper is organized as follows.
In Section~\ref{sec:data}, we introduce our sample and mass measurements.
We calibrate our X-ray hydrostatic masses with lensing in Section~\ref{sec:xlratio} and investigate the robustness of the measurement in Section~\ref{sec:systematics}.
We measure the average concentration for the relaxed cluster sample in Section~\ref{sec:ave_concen}.
Finally, in Section~\ref{sec:discussion}, we provide physical explanations for our results and place them in context with other literature efforts.
Additionally in Section~\ref{sec:discussion}, we discuss our blinding strategy and examine the cosmology dependence of the lensing to X-ray mass ratio.
Concluding remarks are presented in Section~\ref{sec:summary}.

Unless otherwise noted, all mass measurements assume a flat
$\Lambda$CDM reference cosmology with $\Omega_{\rm m} = 0.3$,
$\Omega_{\Lambda} = 0.7$ and $H_0 = 100 \,h\, \mbox{km/s/Mpc}$, where
$h=0.7$.


\section{Observations, Mass Measurements, \& Statistical Methods}
\label{sec:data}

In this section, we describe how the sample was selected and review how the X-ray and lensing masses were measured.

\subsection{Sample Selection}

Our sample selection is motivated by the goal of measuring the lensing to X-ray mass ratio in as many massive, dynamically relaxed clusters as possible, while maintaining robust control of systematic uncertainties.
As described in Paper~I, we conducted a systematic search for relaxed clusters from 361 cluster observations in the \chandra{} and ROSAT archives.
The dynamical state of each cluster was evaluated using an automated morphological classification.
Three discriminators were used to determine the morphological state of a cluster: sharpness of the surface brightness peak, offsets in the centers of neighboring isophotes, and offsets of individual isophotes from a global center.
Surface brightness levels used for the isophote measurements were selected to probe equivalent physical scales, assuming clusters evolve following the self-similar model of Kaiser \citetext{\citealp{kaiser86}; see also \citealp{santos08}}.
This procedure was designed to fairly compare clusters at varying redshifts and signal-to-noise levels, while also avoiding strong assumptions about cosmology (for example, the expansion history).
The employed criteria correlate with traditional measures of morphological disturbance such as surface brightness ``concentration'' and centroid variance (see Paper I for details).
Paper II imposed further requirements for the inclusion of targets in the cosmological analysis, namely a cut on the average gas temperature ($kT \ge 5 \mathrm{keV}$), as well as stricter data quality cuts.

For the lensing comparison, we use the Weighing the Giants (WtG) sample \citep{paper1}, which comprises 51 X-ray selected clusters with Subaru-SuprimeCam follow-up data \citep{mks02}.
The clusters used in this analysis are  listed in Table~\ref{tab:targets}.
We refer to the 12 clusters in common between WtG and the dynamically relaxed cluster sample above  as the ``relaxed WtG'' sample in this paper.
For our tests of morphology dependence of the lensing to X-ray ratio, we supplemented this sample with five additional clusters common to \citet{allen08} and WtG that fail at least one morphological criteria of Paper~I, which we call the ``marginal'' sample.
By using the WtG analysis, we benefit from uniformly derived, accurate WL masses for all clusters in the comparison.

\begin{table*}
  \centering
  \caption[]{%
    The sample of massive, dynamically relaxed clusters used in this work. Column [1] name; [2] adopted redshift; 
[3], [4] J2000 coordinates of adopted cluster center;
[6] \chandra{} clean exposure time (ks); [7] Optical filter coverage; [8] Lensing Band (exposure in s, seeing). See Paper I for details of the X-ray data-set and processing; see \citet{paper1} for details of the lensing data-set, filter definitions, and processing.
  }
  \begin{tabular}{llcc  c  c c }
    \hline
    \textbf{Cluster} & \textbf{z}  & \textbf{RA} & \textbf{Dec} & \textbf{X-ray Exposure} & \textbf{Optical Filters} & \textbf{Lensing Band} \\
\hline
\multicolumn{7}{c}{\textbf{Relaxed WtG Sample}}\\
  Abell 2204       & 0.152      & 16:32:47.1 &  05:34:31.4 & 89.4 & {\it B}$_{\rm J}${\it V}$_{\rm J}${\it R}$_{\rm C}${\it g}$^{\star}${\it r}$^{\star}$ & {\it V}$_{\rm J}$ (1038, 0.58)  \\
  RXJ2129.6+0005   & 0.235      & 21:29:39.9 &  00:05:18.3 & 36.8 & {\it B}$_{\rm J}${\it V}$_{\rm J}${\it R}$_{\rm C}${\it i}$^{+}$ & {\it V}$_{\rm J}$ (1863, 0.58)  \\
  Abell 1835       & 0.252      & 14:01:01.9 &  02:52:39.0 & 205.3 & {\it V}$_{\rm J}${\it I}$_{\rm C}${\it i}$^{+}${\it g}$^{\star}${\it r}$^{\star}$ & {\it i}$^{+}$ (1944, 0.91)  \\
  MS2137.3-2353    & 0.313      & 21:40:15.2 & -23:39:40.0 & 63.2 & {\it B}$_{\rm J}${\it V}$_{\rm J}${\it R}$_{\rm C}${\it I}$_{\rm C}${\it z}$^{+}$ & {\it R}$_{\rm C}$ (5562, 0.57)  \\
  MACSJ1115.8+0129 & 0.355      & 11:15:51.9 &  01:29:54.3 & 45.3 & {\it V}$_{\rm J}${\it R}$_{\rm C}${\it I}$_{\rm C}$ & {\it R}$_{\rm C}$ (1944, 0.65)  \\
  RXJ1532.9+3021   & 0.363      & 15:32:53.8 &  30:20:58.9 & 102.2 & {\it B}$_{\rm J}${\it V}$_{\rm J}${\it R}$_{\rm C}${\it I}$_{\rm C}${\it z}$^{+}${\it u}$^{\star}$ & {\it R}$_{\rm C}$ (2106, 0.55)  \\
  MACSJ1720.3+3536 & 0.391      & 17:20:16.8 &  35:36:27.0 & 53.2 & {\it B}$_{\rm J}${\it V}$_{\rm J}${\it R}$_{\rm C}${\it I}$_{\rm C}${\it z}$^{+}$ & {\it V}$_{\rm J}$ (1944, 0.69)  \\
  MACSJ0429.6-0253 & 0.399      & 04:29:36.1 & -02:53:07.5 & 19.3 & {\it V}$_{\rm J}${\it R}$_{\rm C}${\it I}$_{\rm C}$ & {\it R}$_{\rm C}$ (2592, 0.73)  \\
  RXJ1347.5-1145   & 0.451      & 13:47:30.6 & -11:45:10.0 & 67.3 & {\it B}$_{\rm J}${\it V}$_{\rm J}${\it R}$_{\rm C}${\it I}$_{\rm C}${\it z}$^{+}${\it u}$^{\star}${\it g}$^{\star}${\it r}$^{\star}${\it i}$^{\star}${\it z}$^{\star}$ & {\it R}$_{\rm C}$ (2592, 0.69)  \\
  MACSJ1621.6+3810 & 0.461      & 16:21:24.8 &  38:10:09.0 & 134.0 & {\it B}$_{\rm J}${\it V}$_{\rm J}${\it R}$_{\rm C}${\it I}$_{\rm C}${\it z}$^{+}${\it u}$^{\star}$ & {\it I}$_{\rm C}$ (1568, 0.52)  \\
  MACSJ1427.3+4408 & 0.487      & 14:27:16.2 &  44:07:31.0 & 51.0 & {\it V}$_{\rm J}${\it R}$_{\rm C}${\it z}$^{+}$ & {\it R}$_{\rm C}$ (2544, 0.59)  \\
  MACSJ1423.8+2404 & 0.539      & 14:23:47.9 &  24:04:42.3 & 123.0 & {\it B}$_{\rm J}${\it V}$_{\rm J}${\it R}$_{\rm C}${\it I}$_{\rm C}${\it z}$^{+}${\it u}$^{\star}$ & {\it I}$_{\rm C}$ (1944, 0.73)  \\
\hline
\multicolumn{7}{c}{\textbf{Marginal Sample}}\\
A963 & 0.206 & 10:17:03.562 & 39:02:51.51 & 38.3 &{\it V}$_{\rm J}${\it R}$_{\rm C}${\it I}$_{\rm C}$ & {\it I}$_{\rm C}$ (2700, 0.61) \\
A2390 & 0.233 & 21:53:37.070 & 17:41:45.39 & 79.4 &{\it B}$_{\rm J}${\it V}$_{\rm J}${\it R}$_{\rm C}${\it I}$_{\rm C}${\it i}$^{+}${\it z}$^{+}${\it u}$^{\star}$ & {\it R}$_{\rm C}$ (3420, 0.56)\\
A611 & 0.288 & 08:00:56.818 & 36:03:25.52 & 30 &{\it B}$_{\rm J}${\it V}$_{\rm J}${\it R}$_{\rm C}${\it I}$_{\rm C}${\it g}$^{\star}${\it r}$^{\star}$ & {\it I}$_{\rm C}$ (1896, 0.62) \\
MACSJ0329.6-0211 & 0.450 & 03:29:41.459 & -02:11:45.52 & 22.2 & {\it B}$_{\rm J}${\it V}$_{\rm J}${\it R}$_{\rm C}${\it I}$_{\rm C}${\it z}$^{+}${\it u}$^{\star}$ & {\it V}$_{\rm J}$ (1944, 0.55) \\
MACSJ0744.8+3927 & 0.698 & 07:44:52.310 & 39:27:26.80 & 73.2 &{\it B}$_{\rm J}${\it V}$_{\rm J}${\it R}$_{\rm C}${\it I}$_{\rm C}${\it i}$^{+}${\it z}$^{+}${\it u}$^{\star}$ & {\it R}$_{\rm C}$ (4869, 0.56) \\
\hline
  \end{tabular}
  \label{tab:targets}
\end{table*}


\subsection{X-ray Hydrostatic Masses}

Our X-ray mass estimates are derived from \chandra{} data, using the analysis procedure described in Papers I and II.
However, we have used a newer version of the \chandra{} calibration, resulting in updated hydrostatic masses.
For this paper, the data were processed with CIAO\footnote{\url{http://cxc.harvard.edu/ciao}} version 4.6.1 and CALDB\footnote{\url{http://cxc.harvard.edu/caldb}} version 4.6.2 following standard procedures.\footnote{\url{http://cxc.harvard.edu/ciao/guides/acis_data.html}} 
The intracluster medium was modeled as a spherically symmetric, piece-wise isothermal atmosphere in hydrostatic equilibrium with a gravitational potential given by the Navarro-Frenk-White \citep[NFW; ][]{nfw97} model. 
The center for this deprojection was chosen to maximize the symmetry of the emission at radii $\sim r_{2500}$, where the mass profiles are best constrained (and where we perform the comparison to lensing masses). 
The radius $r_{2500}$ is where the mean enclosed density is 2500 times the critical density of the universe at the cluster's redshift, as determined by the X-ray observations.
Appropriate foreground and background components were included in the spectral analysis, along with a model for the cluster emission.
The model was fitted to the photon counts in the 0.6-7.0 keV band using the modified C-statistic in XSPEC \citep{xspec} in energy bins, defined to have at least one count per bin.  
See Paper~II for more details.

\subsection{Weak Lensing Masses}
\label{sec:lensing_masses}

The lensing analysis for this work is similar to that presented in the Weighing the Giants papers \citep{paper1, paper2, paper3}.
Here, we summarize the essential elements of the mass measurement process that are referenced later in the text, and highlight all small changes in the modeling procedure.

WL masses were derived from multi-filter observations with SuprimeCam at Subaru and Megacam at CFHT. 
We generated shear catalogs for each cluster field using the KSB$+$ \citep{ksb95, hfk98} moments-based code \textsc{analyseldac}, described in \citet{erben01}.
Absolute shear calibration was determined from the STEP2 simulations, which were designed to replicate SuprimeCam observations \citep{step2}.
The cluster shear profiles were modeled as NFW halos over a cluster-centric radial range 750\,kpc--3\,Mpc.
We adopted the X-ray center when constructing the 1D average shear profiles.
Note that the restriction of shear measurements to relatively large cluster radii makes our mass estimates insensitive to the exact choice of center \citep{paper1}.

Two different methods were used to measure the redshift distribution of galaxies selected as lensed sources: the ``color-cut'' method and the \pz method.
Both methods are described in \citet{paper3}, where we demonstrated that the \pz method is unbiased and that our implementation of the ``color-cut'' method yields consistent results. 
We briefly describe each method below. 
For the analysis of Section~\ref{sec:xlratio}, we adopted the color-cut masses cross-calibrated with \pz masses as our baseline, thus maximizing the sample size (12 clusters versus 6 with the \pz method) while maintaining uniform systematic uncertainties.

For the  ``color-cut'' method, we matched the galaxy population of each cluster field to the COSMOS deep field \citep{ilbert09}, for which high quality 30+ filter photometric redshifts are available.
To combat signal dilution by contaminating cluster members, we removed the red sequence for each cluster.
We then performed a ``contamination correction'', where the measured signal was boosted using the average excess galaxy number counts observed in the WtG sample.

For the \pz method, we calculated photometric redshift probability functions for individual galaxies from five filter {\it B}$_{\rm J}${\it V}$_{\rm J}${\it R}$_{\rm C}${\it I}$_{\rm C}${\it z}$^{+}$ photometry \citep{paper2}.
We measured the NFW halo properties of the cluster with an unbinned Bayesian model fitted to the measured shear at each galaxy position.
During the fit, we marginalized over the redshift probability function for each galaxy.
\pz masses are used in this work to test the robustness of the lensing to X-ray mass ratio measurement and to measure the average concentration of the sample.

We adopted a prior on the  NFW concentration from \citet{neto07}, using results for relaxed clusters from their largest mass bin, $\log_{10} M_{200}h^{-1}\msun = 14.875$--$15.125$.
The prior probability $P(\log_{10} c)$  was modelled as a Gaussian distribution with mean 0.664 and standard deviation 0.061.
Note that this differs slightly from \citet{paper3}, for which the analysis was not restricted to dynamically relaxed clusters, and where a Gaussian prior of $log_{10} c = 0.6 \pm 0.116$ representing clusters in all dynamic states, was used.

Previous work has shown that modelling a sample of cluster lensing observations as NFW halos following a fixed mass-concentration relation can lead to small biases in the average mass of the sample \citep{becker11, bahe12}.
This mass bias can be traced to the use of an incorrect mass-concentration relation and to deviations of the cluster density profile from NFW at radii beyond the cluster virial radius.
To correct both of these effects, we replicate our mass modeling procedure on mock lensing observations of the Millennium-XXL simulation \citep[S. Hilbert et al.~in prep; see also][]{angulo2012}.
Millennium-XXL features 303 billion particles in a 4.1 Gpc box, providing over one hundred halos in a mass range matching the clusters in this study.
We replicated our analysis on two simulation snapshots where mock lensing observations were available, $z=0.25$ and $z=1.0$.

For this study, we find that a correction factor constant in mass is sufficient to characterize the bias in $M_{2500}$.
At $z=0.25$, we find a correction factor of $1.03 \pm 0.01$ (stat only), increasing our masses by $\approx 3\%$.
At $z=1.0$, we find a correction factor of $1.10 \pm 0.02$.
We linearly interpolate the correction to each cluster redshift in our sample and marginalize over the statistical uncertainty when fitting for the lensing to X-ray mass ratio.
We discuss additional systematics associated with this simulation-based correction in Section~\ref{sec:concentration}.
A complete discussion of how we derive this correction and the related systematic error estimates  will be presented in Applegate et al., in prep.
Note that this simulation correction is new since the publication of Paper II in the series.

In \citet{paper3}, we discussed the relevant systematic uncertainties in measuring the mean lensing mass of the full 51 cluster sample.
We attributed a 4\% systematic uncertainty to our shear measurements.
The determination of the photometric-redshift distribution contributed an additional 3\% systematic uncertainty.
We estimated a 4\% uncertainty in the calibration of the color-cut lensing masses by \pz masses.
The total systematic from these three sources is $\approx 6\%$.
We had previously included a 3\% systematic uncertainty attributed to modeling the mass distribution in the clusters.
We update the mass model uncertainty estimate, in light of the new simulation-based correction, in Section~\ref{sec:concentration}.

Triaxiality of cluster halos and line-of-sight structure contribute additional noise to lensing measurements, estimated to be $\approx 20\%$ \citep{hoe03, becker11}.
The Millennium-XXL mock observations allow us to estimate the intrinsic scatter from the matter distribution within 100 h$^{-1}$Mpc (physical) of the cluster center.
We model the intrinsic scatter in the simulations as a log-normal distribution, finding $\sigma_{\mathrm{int}} = 0.23 \pm 0.01$ at $z=0.25$ and $\sigma_{\mathrm{int}} = 0.28 \pm 0.01$ at $z=1.0$.
However, a 100 h$^{-1}$Mpc integrated length does not account for all scatter induced by line of sight structure.
While we could follow the method of \citet{hoe03} to estimate the scatter component induced  by large scale structure, we elect to simply fit for the total magnitude of intrinsic scatter when modeling the lensing to X-ray mass ratio.


\subsection{Statistical Methods}
\label{sec:statmethods}

The X-ray to lensing mass ratio is poorly determined by a simple averaging procedure. 
First, lensing masses generically have large uncertainties that are non-Gaussian (nor log-normal) in shape, even for the most massive clusters with the best data. 
Second, the uncertainties in the lensing masses are correlated with the X-ray masses when measured within an aperture determined from the X-ray data. 
Neglecting this correlation will artificially boost the apparent precision of the measurement.
Third, we expect an approximately log-normal intrinsic scatter between X-ray and lensing masses, since the lensing masses are sensitive to cluster triaxiality and line of sight structure \citep{becker11, bahe12}. 
This scatter is of a comparable magnitude to the WL measurement uncertainties, so neither can be safely neglected.

While none of these issues are new, their combination limits the applicability of standard approaches. 
Fitting packages such as \textsc{linmix\_err} from \citet{kelly07} or \textsc{BCES} from \citet{bces} assume that the forms of the measurement error and the intrinsic scatter are the same (Gaussian). 
Both codes also restrict users to a simple two-parameter linear model.

To avoid these pitfalls, we can instead write down the posterior probability distribution for the lensing to X-ray mass ratio without approximation. 
Specifically, we can incorporate the exact probability distributions for NFW halo parameters that describe the lensing and X-ray data sets. 
This correctly incorporates both detections and non-detections into the ratio, while also allowing us to directly account for the correlation between lensing and X-ray masses. 
We model the intrinsic scatter from triaxiality and line of sight structure as a log-normal distribution, independent of the form of the measurement errors. 
Appendix~\ref{sec:likelihood} explicitly presents the posterior probability function for the lensing to X-ray mass ratio. 

This basic ratio model can be easily extended to generalize the relationship between lensing and X-ray masses or to linearly regress the ratio against other variables.
When the independent variables in these fits have non-negligible uncertainty, for example the cluster morphology discriminators we consider in Section~\ref{sec:morphology}, we must also include a prior on the true, unobserved distribution of that variable.
For this task, we adopt the hierarchical mixture of Gaussians prior employed in \citet{kelly07}, with uniform priors on Gaussian scale parameters (\ie $\sigma$) following \citet{gelman2006}.
By default, we report results for mixtures of two Gaussians, but have verified that results are not sensitive to the exact number employed.

Throughout the paper, we adopt uninformative priors when possible.
For example, we adopt a uniform prior on the log of lensing to X-ray mass ratio, which equally prefers ratios greater or less than one.
We also adopt uniform priors on intrinsic scatter $\sigma_{\mathrm{int}}$.
We quote maximum a posteriori parameter point estimates,  transforming probability distributions and point estimators appropriately when log transforms are applied to variables.

For the majority of this paper, we consider the lensing to X-ray mass ratio at a fixed cosmology. 
However, both the lensing and X-ray masses inferred from the data implicitly depend on the cosmological model.
We comment on the resulting dependence of the lensing to X-ray mass ratio on cosmological parameters in Section~\ref{sec:cosmo_dependence}.
When using the lensing data to calibrate cluster masses in a cosmological test, such as the gas-mass fraction (\fgas) constraints of Paper~II, it is best to directly incorporate the full model described above; for each trial cosmology, model parameters are inferred from the measured shear profiles (as a function of angle) and galaxy redshift distributions. 
In the case of Paper~II (see also \citealt{allen08}), the model for the X-ray data includes a parameterized redshift-dependent scaling, $K(z)=K_0(1+K_1 \,z)$, encoding the average ratio of true cluster masses to X-ray mass estimates at a given cluster redshift. 
Given the expectation that our lensing masses are unbiased within a 7\% tolerance, we can interpret this function as the lensing to X-ray mass ratio, and use the likelihood associated with the lensing data (Appendix~\ref{sec:likelihood}) to constrain the model for $K(z)$ (in practice, the current data can constrain $K_0$, but $K_1$ must be constrained by a prior). 
For simplicity, the implementation in Paper~II neglects the cosmology dependence of the mass--concentration relation as well as the sub-dominant uncertainties from the contamination correction and the lensed galaxy redshift distribution (characterized by $\langle\beta\rangle$ and $\langle\beta^2\rangle$); due to the small size of the comparison sample, these effects are much smaller than the statistical uncertainties.
The same approach was also used in \citet{paper4}.


\section{Weak Lensing to Chandra X-ray Ratio}
\label{sec:xlratio}

In this section we present our measurements of the lensing to \chandra{} X-ray mass ratio for our sample of 12 relaxed WtG clusters.
We first present results for the ratio measured within $r_{2500}$, where this radius is determined from the X-ray data.
This is the appropriate ratio to use in the \fgas model described in Paper~II.
We then present results for the mass ratio $M_{2500}^L/M_{2500}^X$, where each data set independently determines its own value of $r_{2500}$.
Finally, we also report the WL to X-ray mass ratio measured within the X-ray defined $r_{500}$.

\begin{figure}
\includegraphics[width=\columnwidth]{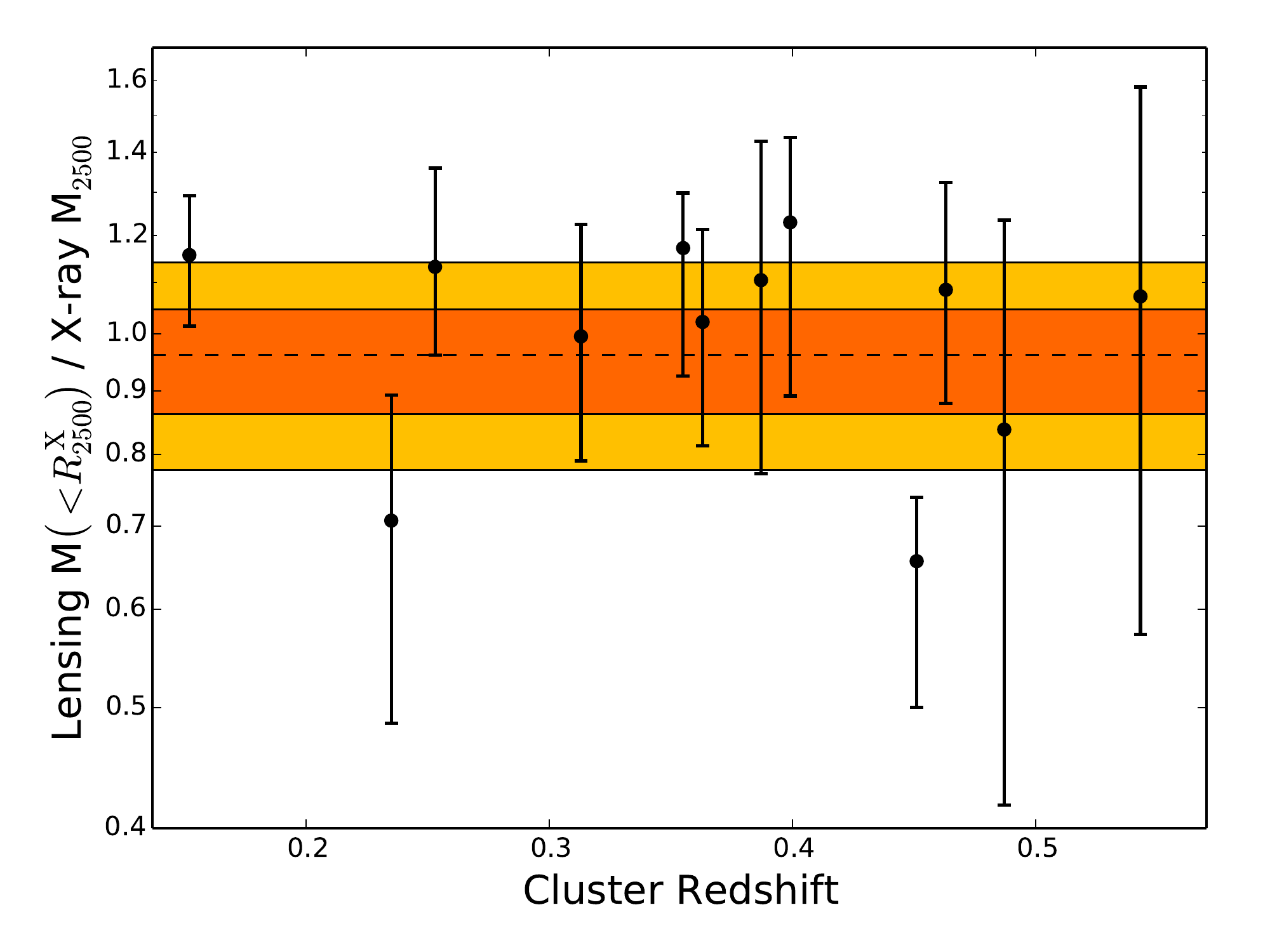}
\caption{Measured weak lensing to \chandra{} X-ray mass ratio plotted against cluster redshift for the relaxed WtG sample of clusters. Lensing masses are measured within the X-ray defined \rmeas{} and use the color-cut analysis. Data points are for individual clusters, where error bars represent 68\% confidence intervals. The dashed line and orange bands indicate the ensemble mass ratio, 68\% and 95\% ranges. The ratio and statistical 68\% range is $0.967_{-0.092}^{+0.063}$, or $0.956 \pm 0.082$ when approximated as a Gaussian distribution. Additional systematic uncertainties of 9\% apply, for a total uncertainty of $\approx 12\%$. The y-axis and associated probability distributions are  log transformed.}
\label{fig:ratio_v_redshift}
\end{figure}

We first consider the mass ratio within the X-ray defined \rmeas, as plotted in Figure~\ref{fig:ratio_v_redshift}.
We measure a mass ratio of $M_L/M_X = 0.967_{-0.092}^{+0.063}$, where the quoted uncertainties are the statistical 68\% confidence interval after marginalizing over the intrinsic scatter. 
The best Gaussian approximation of the posterior PDF is $M_L/M_X = 0.956 \pm 0.082$.
From the statistical uncertainties alone, the ratio is fully consistent with unity.

The measurement is also subject to systematic uncertainties from the WtG WL analysis (Section~\ref{sec:lensing_masses}).
Whereas the  WtG analysis of \citet{paper3} measured masses within 1.5\,Mpc, the current measurement is at \rmeas, typically a factor of $\sim3$ smaller.
This requires us to extrapolate our shear measurements inwards, thereby increasing our sensitivity to priors on the mass--concentration relation.
At this radius, the mass--concentration relation contributes an additional 6\% systematic uncertainty, which we discuss in Section~\ref{sec:concentration}.
Therefore, the total WL systematic uncertainty is 9\%, resulting in a combined statistical and systematic uncertainty for the mass ratio measurement of 12\%.

We measure an intrinsic scatter of $\sigma_{\mathrm{int}} = 0.146_{-0.068}^{+0.097}$ present in the lensing to X-ray ratio.
This amount of intrinsic scatter is consistent at $1\sigma$ with the intrinsic scatter predicted by the MXXL simulations and with the intrinsic scatter detected for the full WtG cluster sample \citep{paper4}.


Next, we measure the ratio within $r_{2500}$ apertures determined independently from the lensing and X-ray data.
This alternative ratio is relevant for calibrating mass proxies in some circumstances.\footnote{We reiterate, however, that the result applies only for our reference cosmology, and that in applications where cosmological parameters are allowed to vary, the only robust approach is to fit the mass ratio model simultaneously with the cosmology.}
These results should be noisier, as WL cannot easily measure $r_{2500}$.
The median mass ratio and 68\% statistical uncertainty is $0.90^{+0.12}_{-0.12}$, with a measurement of intrinsic scatter of $\sigma_{\mathrm{int}} = 0.25_{-0.11}^{+0.15}$
The same 9\% systematic uncertainty applies, resulting in a total measurement precision of 16\%.

Finally, we measure the ratio within the X-ray defined $r_{500}$.
Note that this is in some cases an extrapolation of the X-ray data to larger radii, but is well measured by the WL observations.
We measure a WL to X-ray ratio of $1.059_{-0.096}^{+0.092}$ when using a WL correction factor derived from the Millennium-XXL simulations that is appropriate for $r_{500}$.
Again, a total WL systematic uncertainty of 9\% applies.


\section{Discussion of Systematics}
\label{sec:systematics}

We investigate the possible influence of a number of systematics that in principle could alter the measured mass ratio.
We start with the adopted \chandra{} temperature calibration.
We then verify the robustness of the measurement to the definition of the cluster sample and the details of the lensing analysis.
Finally, we search for trends in the X-ray to lensing ratio that correlate with cluster mass or redshift.

\subsection{Chandra Calibration}
\label{sec:calib_dependence}

X-ray hydrostatic masses, and therefore the lensing to X-ray mass ratio, are dependent on the effective area calibration adopted for \chandra.
While \chandra{} calibration updates usually refine the most recent observations, some retroactively affect temperatures  measured from older data.
One such update happened after the publication of Paper II of this series.
To see how large an effect these calibration changes have on our measurements, we compared our measurements made above (using CIAO version 4.6.1 and CALDB version 4.6.2) with the calibration from May 2012 (CALDB version 4.4.10 and CIAO version 4.4) used in Paper~II.

When X-ray mass estimates were compared between the two calibration versions for the 40 clusters in the full relaxed sample from Paper~I, we saw that the newer calibration (the default used in this paper) lowers the average X-ray mass, $M_{2500}$, by $5\%$, with a scatter of $5\%$. 
Restricting to only the 12 relaxed WtG clusters in the calibration sample examined in this paper, we see a downward shift of only $\approx3\%$.
The small shift in X-ray masses is reflected in an equally small shift in the lensing to X-ray mass ratio. 
With the May 2012 calibration, the lensing to X-ray mass ratio at the X-ray measured $r_{2500}$ is $0.940_{-0.081}^{+0.065}$, in comparison to our baseline results of $0.967_{-0.092}^{+0.063}$.

This shows that our results are not overly sensitive to recent changes in the \chandra{} effective area calibration.
More importantly, the use of lensing data to calibrate the X-ray masses in cosmological tests like that of Paper~II corrects for any overall mass bias potentially introduced by the \chandra{} calibration.


\subsection{Sample Selection}
\label{sec:morphology}

In Papers I and II, clusters were included in the cosmological analysis with the \fgas{} test based on X-ray morphology indicators.
Any detected trend with morphology would indicate that the lensing to X-ray mass ratio, and therefore the \fgas test, is sensitive to the selection criteria.
In this section, we verify that no trend exists with any of these indicators.

Since by construction all clusters in the relaxed WtG sample qualify as relaxed, we extended our sample with five additional clusters that failed one or more morphology criteria in Paper~I (the ``marginal'' sample).
Each of the five clusters have previously been included in subjective selections of relaxed clusters \citep{allen08, donahue14}.

As a first check, we compared the measured lensing to X-ray mass ratio for the relaxed WtG sample against the marginal sample.
Using the five clusters in the marginal sample alone, the measured ratio within the X-ray measured \rmeas{} is $M_L/M_X = 1.05^{+0.24}_{-0.23}$.
In addition, we detect an intrinsic scatter of $\sigma_{\mathrm{int}} = 0.44^{+0.25}_{-0.21}$.
The ratio is consistent with the twelve clusters from the relaxed WtG sample, albeit with low precision.
When all seventeen clusters are considered, the ratio is $M_L/M_X = 0.979_{-0.076}^{+0.078}$ with an intrinsic scatter of $\sigma_{\mathrm{int}} = 0.22_{-0.075}^{+0.080}$, again consistent with the results of the relaxed-only sample.

\begin{figure*}
  \centering
  \begin{minipage}{0.32\linewidth}
    \centering
    \includegraphics[width=\columnwidth]{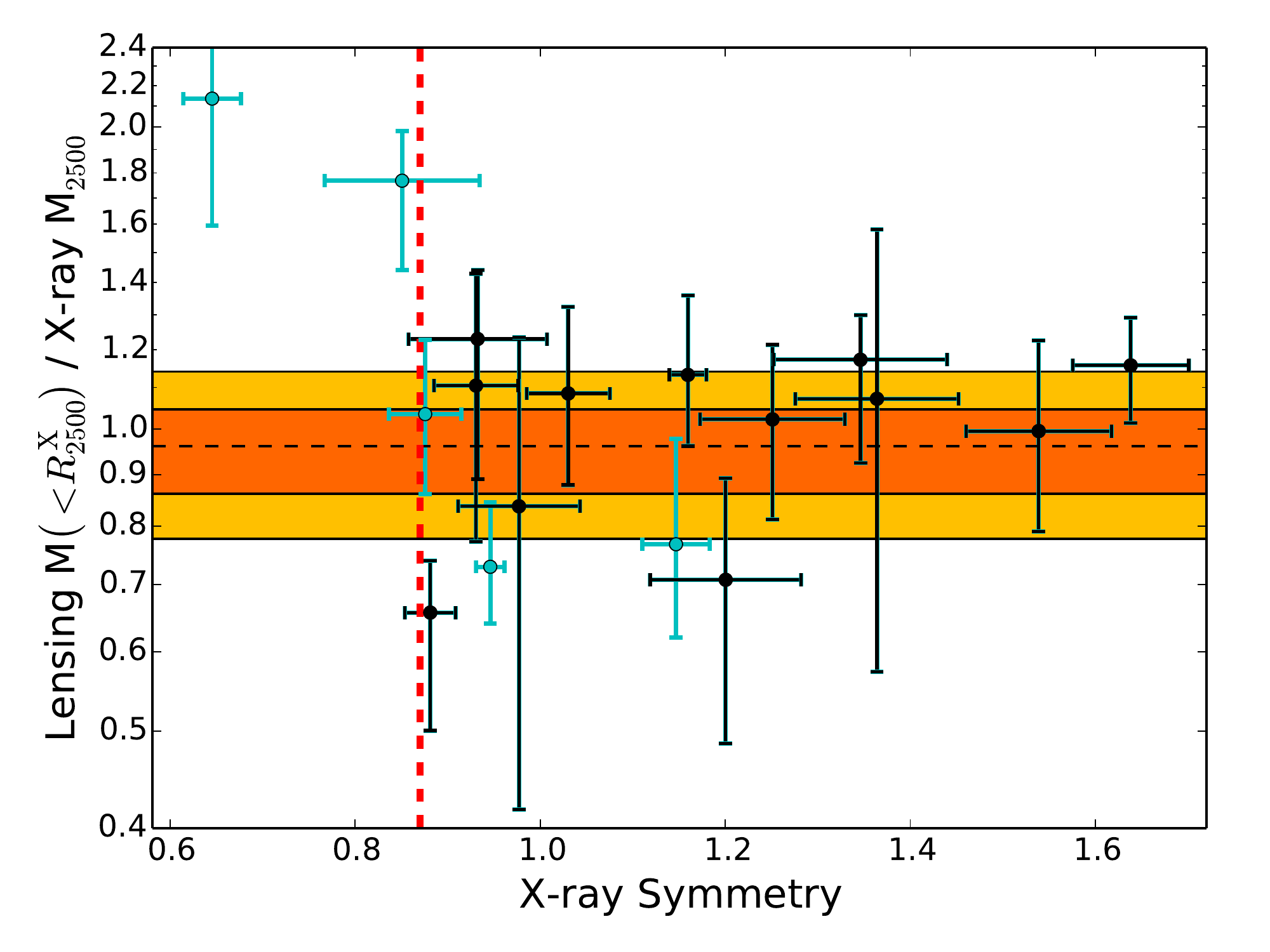}
  \end{minipage}
  \begin{minipage}{0.32\linewidth}
    \centering
    \includegraphics[width=\columnwidth]{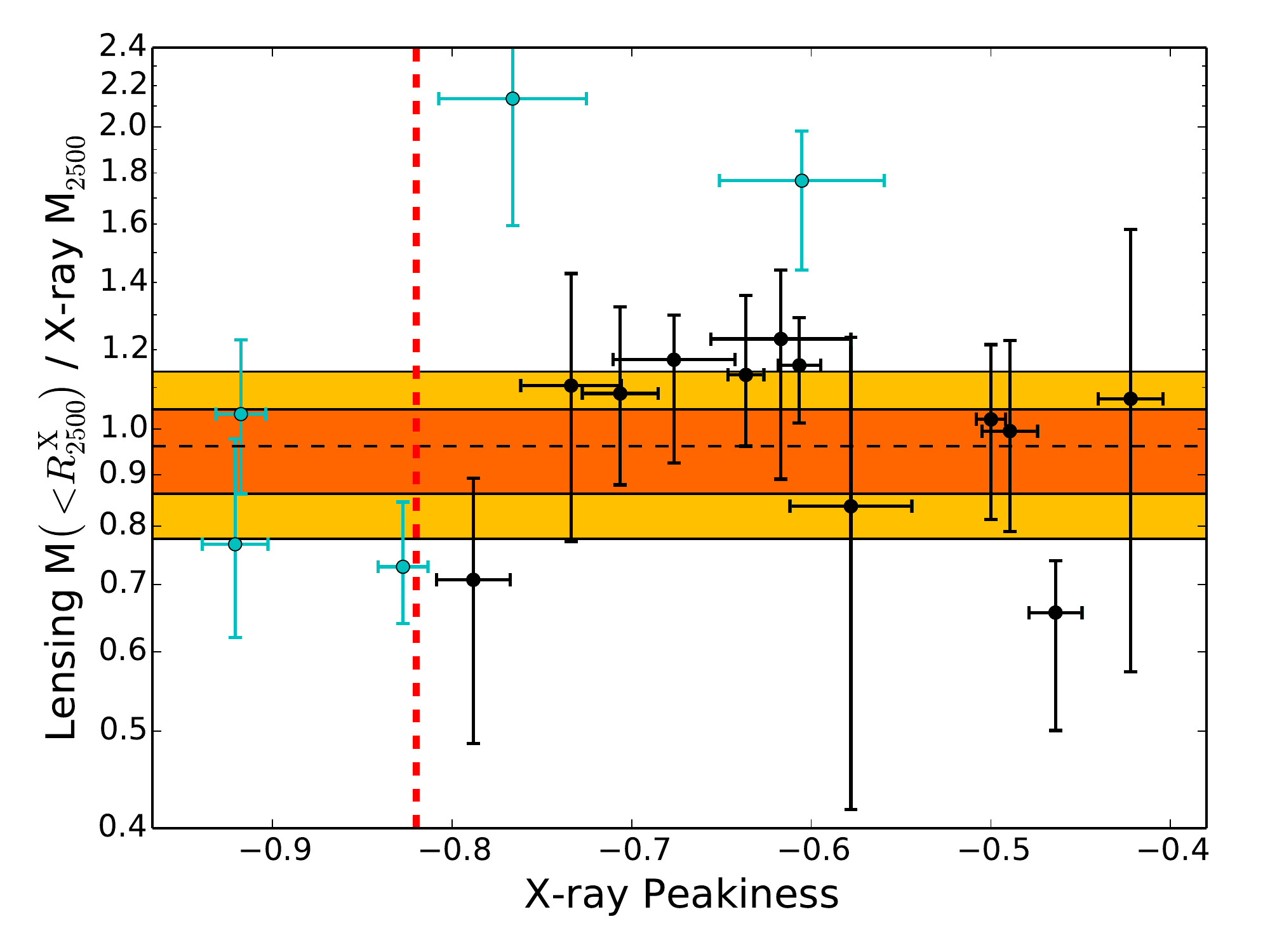}
  \end{minipage}
  \begin{minipage}{0.32\linewidth}
    \centering
    \includegraphics[width=\columnwidth]{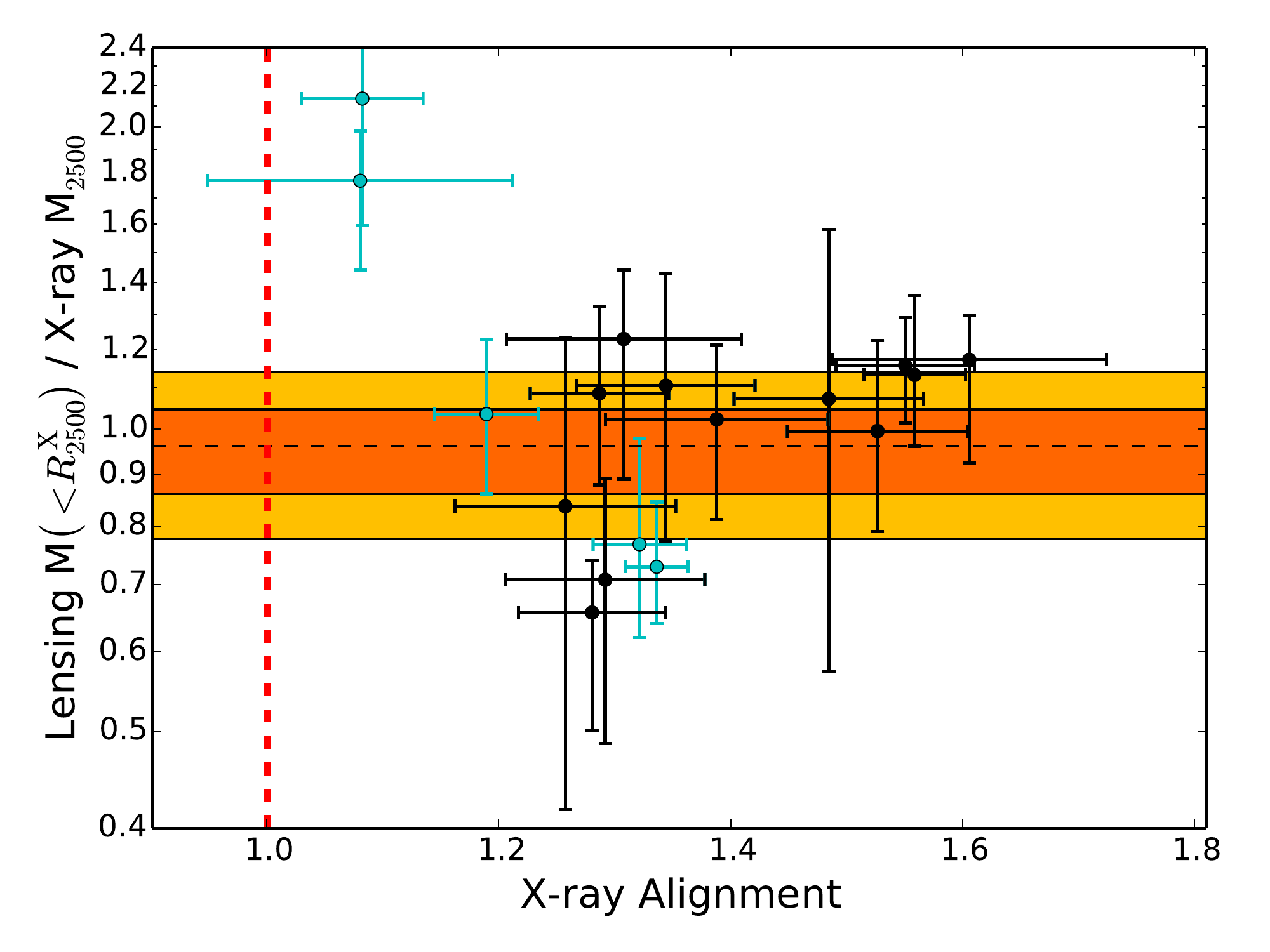}
  \end{minipage}
  \caption{The sample ratio (black dotted), 68\% and 95\% confidence intervals (orange, shaded),  and ratios for individual clusters, plotted against three X-ray morphology measurements: (left) symmetry, (center) peakiness, and (right) alignment. The vertical dashed red lines show the thresholds that define the relaxed region. Black data points are the relaxed WtG sample, while blue data points are the marginal sample.}
  \label{fig:morphology}
\end{figure*}

\begin{table*}
  \centering
  \caption[]{Measured slopes and offsets for the best fitting linear relationship between the lensing to X-ray mass ratio versus three morphology measures, for the relaxed WtG and marginal samples. Quoted are the 1D 68\% marginalized uncertainties for each quantity.}
  \begin{tabular}{l | c c c | c c c}
    \hline
     & \multicolumn{3}{ c |}{\textbf{Relaxed WtG Sample}} & \multicolumn{3}{ c }{\textbf{Relaxed WtG $+$ Marginal Sample}} \\
    \textbf{Morphology Measure} & Slope & Offset & Pivot & Slope & Offset & Pivot \\
    \hline
    Symmetry     & $0.42_{-0.24}^{+0.37}$    & $0.94_{-0.06}^{+0.07}$ & 1.18   & $0.07_{-0.45}^{+0.25}$  & $0.97_{-0.08}^{+0.08}$ & 1.10  \\
    Peakiness    & $-0.74_{-0.53}^{+1.02}$   & $0.95_{-0.08}^{+0.08}$ & -0.602 & $0.04_{-0.53}^{+0.61}$  & $0.99_{-0.09}^{+0.07}$ & -0.66\\
    Alignment    & $1.30_{-0.80}^{+0.96}$ & $0.95_{-0.10}^{+0.06}$ & 1.41      & $-0.40_{-0.57}^{+1.01}$ & $0.97_{-0.07}^{+0.08}$ & 1.35 \\

    \hline
  \end{tabular}
\label{table:morphology_fits}
\end{table*}

We also investigated linear trends in the lensing to X-ray mass ratio with each of the morphology measures used to define the sample (see Paper I).
The mass ratio is plotted against symmetry, peakiness, and alignment in figure~\ref{fig:morphology}. 
To investigate trends in these parameters, we adopt a linear model between each parameter and the log mass ratio, $\ln M_{2500}^L/M_{2500}^X = \alpha + \beta X$ (see Section~\ref{sec:statmethods}).

We list the 1D marginalized uncertainties for the slope and intrinsic scatter for each morphology measure and both samples in table~\ref{table:morphology_fits}.
When considering the 12 clusters in our relaxed WtG sample, both Symmetry and Alignment exhibit an $\approx 1\sigma$ preference for a non-zero slope.
However, no trend  is evident when the combined ``relaxed WtG'' and ``marginal'' samples are considered.
We note that the intrinsic scatter appears to increase for clusters closer to the multidimensional selection boundary, as seen in Figure~\ref{fig:morphology}.

From these tests, we conclude that our results on the WL to X-ray mass ratio are not overly sensitive to the exact morphological selection criteria used to select the sample.


\subsection{Lensing Analysis}

Weighing the Giants lensing masses measured within an aperture of at $1.5$ Mpc should be unbiased within a 7\% systematic uncertainty \citep{paper3}.
However, since the measurements here are at the smaller radius of $\approx r_{2500}$ and employ a new simulation-based correction factor, the systematic uncertainties of two components of the lensing analysis are worth reviewing.
First, we explore how the lensing to X-ray mass ratio varies when we change how we calculate the background redshift distribution of lensed sources and correct for cluster galaxy contamination.
Then, we quantify the systematic uncertainty from the mass model and the simulation correction.


\subsubsection{Lensed Source Redshift Distribution}

The Weighing the Giants project employed two methods to estimate the redshift distribution of lensed sources and address signal dilution by cluster galaxies, the color-cut method and the \pz method.
The color-cut method statistically matches the galaxy population in the cluster field with a reference deep field and corrects the shear profile for the diluting effects of cluster galaxies. 
In contrast, the \pz method estimates a redshift probability function for each galaxy in the cluster field and only includes galaxies that are believed to be in the background of the galaxy cluster.
We used the color-cut method in this study to maximize our sample size while maintaining uniform systematic uncertainties.
We previously showed that our mass measurements using the \pz method were consistent with our implementation of the color-cut method used in this analysis \citep{paper3}.
Here we check if using the alternative \pz method produces consistent results for the six clusters in the relaxed WtG sample where a comparison is possible.

\begin{figure}
\centering
\includegraphics[width=\columnwidth]{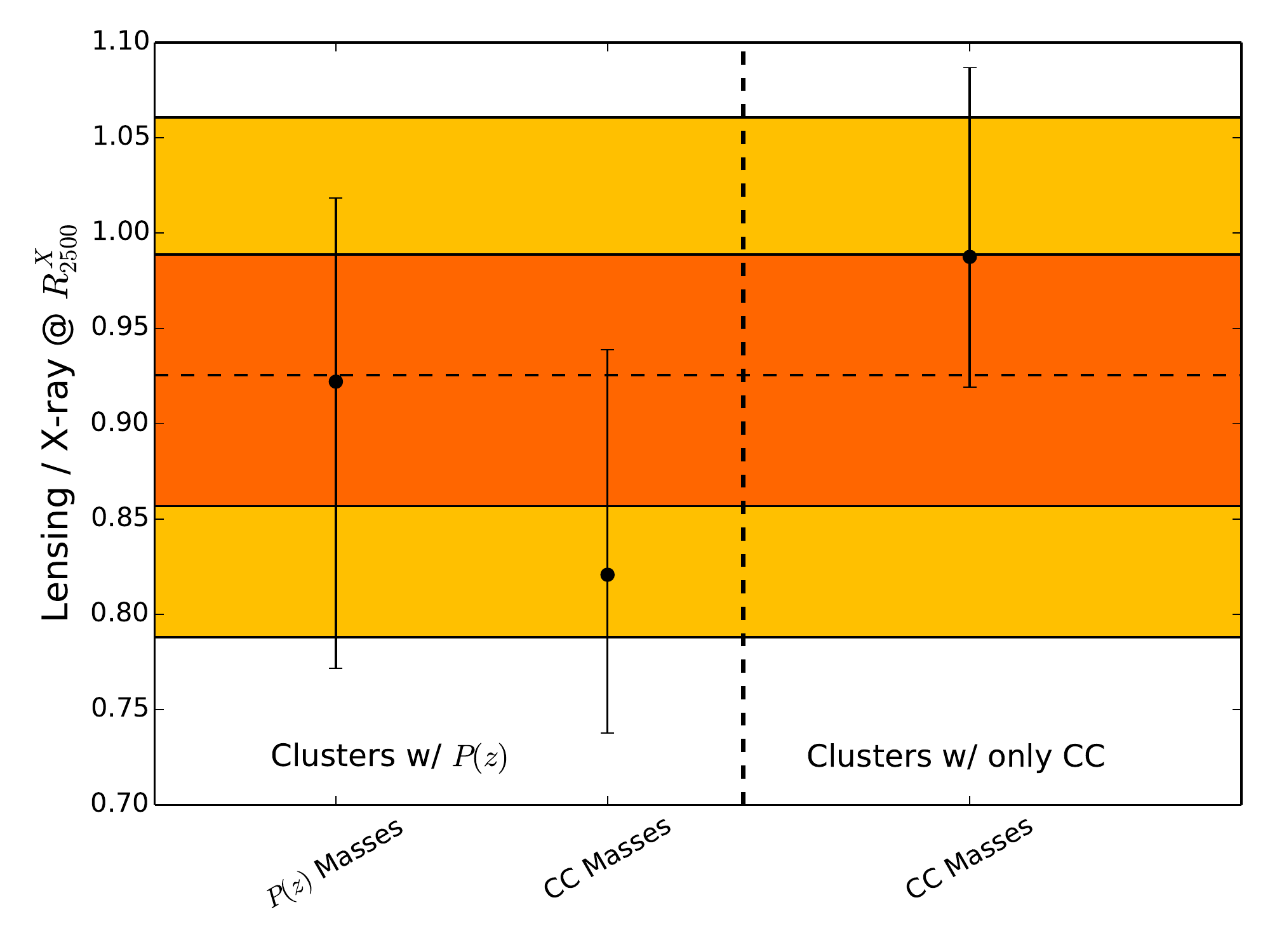}
\caption{Results for three subsets of six clusters, using both the color-cut analysis and the \pz analysis. Data points with error bars show 68\% confidence intervals for each cluster subset, while the dark orange and light yellow shaded regions are the 68\% and 95\% confidence intervals for the full sample. Groups one and two are identical clusters, but analyzed using the \pz and color-cut methods, respectively. Group 3 comprises six independent clusters where only the color-cut analysis can be applied. Groups 1 and 2 are statistically consistent with group 3 at the $1\sigma$ level. Groups 1 and 2 are also statistically consistent when the covariance between the two measurement methods is fully accounted for. Note that the correction for biases in the mass model has not been applied here. }
\label{fig:cluster6}
\end{figure}

Figure~\ref{fig:cluster6} shows the constraints on the lensing to X-ray mass ratio within the X-ray measured $r_{2500}$ for the two sets of clusters in the analysis, those with and without \pz mass measurements.
For the clusters with \pz mass measurements, we calculated the lensing to X-ray mass ratio twice, once with the \pz method and once with the color-cut method.
Both analyses are statistically consistent with the independent set of color-cut only clusters.

For the six clusters where both the color-cut and \pz methods can be applied, the measured ratios differ by 11\%.
Scatter between the methods is expected: each method selects a somewhat different population of lensed galaxies for analysis, resulting in $\approx20\%$ scatter between these measurements at $z \approx 0.2$, growing to $\approx40\%$ at $z\approx 0.6$ \citep{paper3}.
To evaluate the statistical significance of the observed offset between methods for the six clusters in our sample, we created random samples of six clusters from the set of 27 clusters with \pz measurements in \citet{paper3}.
The distribution of offsets between \pz method masses and color-cut method masses for these subsamples is consistent with zero and has a population standard deviation of $\sigma=16\%$.
The offset detected between analyses for these six clusters is therefore within expectations.

\subsubsection{Lensing Mass Model}
\label{sec:concentration}

For this study, we have used the NFW density profile to model the observed shear signal of galaxy clusters.
We have also assumed a mass-concentration relation, both to reduce the nonlinearity of the model and to compensate for the inability of the lensing data to constrain the density profile shape.
However, studies of mock lensing observations of N-body simulations have shown that modeling a population of realistic clusters as isolated NFW halos can lead to small biases in the average mass of the sample \citep{becker11, bahe12}.
Lensing masses may also be sensitive to the choice of mass-concentration relation, either when modeling shear information at small cluster-centric radii or, as in this work, when extrapolating mass measurements to small radii.

To correct for these effects, we replicated our lensing measurement procedure on mock observations of the Millennium-XXL (MXXL) N-body simulation and introduced a correction factor (see Section~\ref{sec:lensing_masses}).
While the statistical uncertainty of the correction factor is small (less than 2\%), additional systematic uncertainties arise from mismatches between the simulation and reality.
The mass-concentration relation introduces the largest systematic uncertainty.
The simulation correction factor effectively fixes our choice of mass-concentration relation, on average altering the masses to what we would have measured using the mass-concentration relation realized in the MXXL. 
However, a residual bias may remain due to the cosmology dependence of the mass-concentration relation \citep{ludlow2013, diemer15}.
We quantify our sensitivity to the mass-concentration relation by varying the peak of our prior on concentration by $\pm 20\%$.
This range is motivated by \citet{bhattacharya13}, who measured differences in the mass-concentration relation amplitude of $\approx 15\%$ between their results and \citet{duffy08}, as well as between results based on the Millennium Simulation \citep{neto07, gao08, hayashi08}.
The measured WL to X-ray mass ratio at $r_{2500}$ varied by $\pm 6\%$.

The simulation correction derived from MXXL is based on a mass-limited selection of halos.
In contrast, the morphology selection from Paper I should select clusters that have less substructure than the general population, yet has not significantly altered the ellipticity distribution of the clusters \citet{mantz_morph}.
This should on average decrease the magnitude of observed intrinsic scatter but not alter the correction factor \citep{bahe12}.
Based on the measured ellipticity distribution of the relaxed sample from Paper I, we also consider it unlikely that a conspiracy exists between a preferential selection of clusters projected along the line of sight, leading to overestimated lensing masses, and a biased Chandra temperature calibration, leading to overestimated X-ray masses.
Future work is required to replicate the Paper I selection function in cosmological simulations.

In principle, lensing mass measurements can be sensitive to the choice of center used to measure the shear profile.
However, we have previously shown that our shear measurements are minimally affected by the choice of cluster center \citep{paper1}.
We have also studied the effects of shear profile miscentering by introducing random offsets into the MXXL mock lensing observations.
When we perturb the assumed center following a probability distribution of BCG-X-ray offsets observed in Chandra archive observations of clusters (Paper I), we see a sensitivity in the correction factor of only $\approx 3\%$.
In addition, the relaxed WtG sample shows considerably reduced BCG-X-ray offsets than the general population, a further indication that these clusters are indeed dynamically relaxed (Paper I).
We therefore expect systematics from miscentering to be negligible for this sample.

We have used two simulated redshifts from the MXXL to measure our correction factor, at $z=0.25$ and $z=1.0$, and have linearly interpolated in redshift between the measurements.
However, the bulk of clusters in the sample are at low redshift.
To test the sensitivity of our measurements to the details of the interpolation scheme, we remeasured the lensing to X-ray ratio using only the correction factor from the $z=0.25$ simulation.
In this case, we measure the ratio to be $0.940_{-0.074}^{+0.081}$, where the best Gaussian approximation to the posterior PDF is $\mu = 0.948 \pm 0.082$, a shift of less than $\approx2\%$.


\subsection{Mass or Redshift Trend}
\label{sec:extensions}


In this section, we search for trends in the X-ray to lensing ratio with mass or redshift.
Such trends would suggest that a more complex calibration model is needed for the \fgas experiment.
Previous work has suggested that a trend with mass may exist for hydrostatic masses from \chandra, albeit at low statistical significance and for samples including all dynamic states \citep{israel14}.
In our case, we emphasize that the baseline ratio model is an adequate description of the data, and that the small size of our sample makes precise measurements of extended models difficult.

\begin{figure}
\includegraphics[width=\columnwidth]{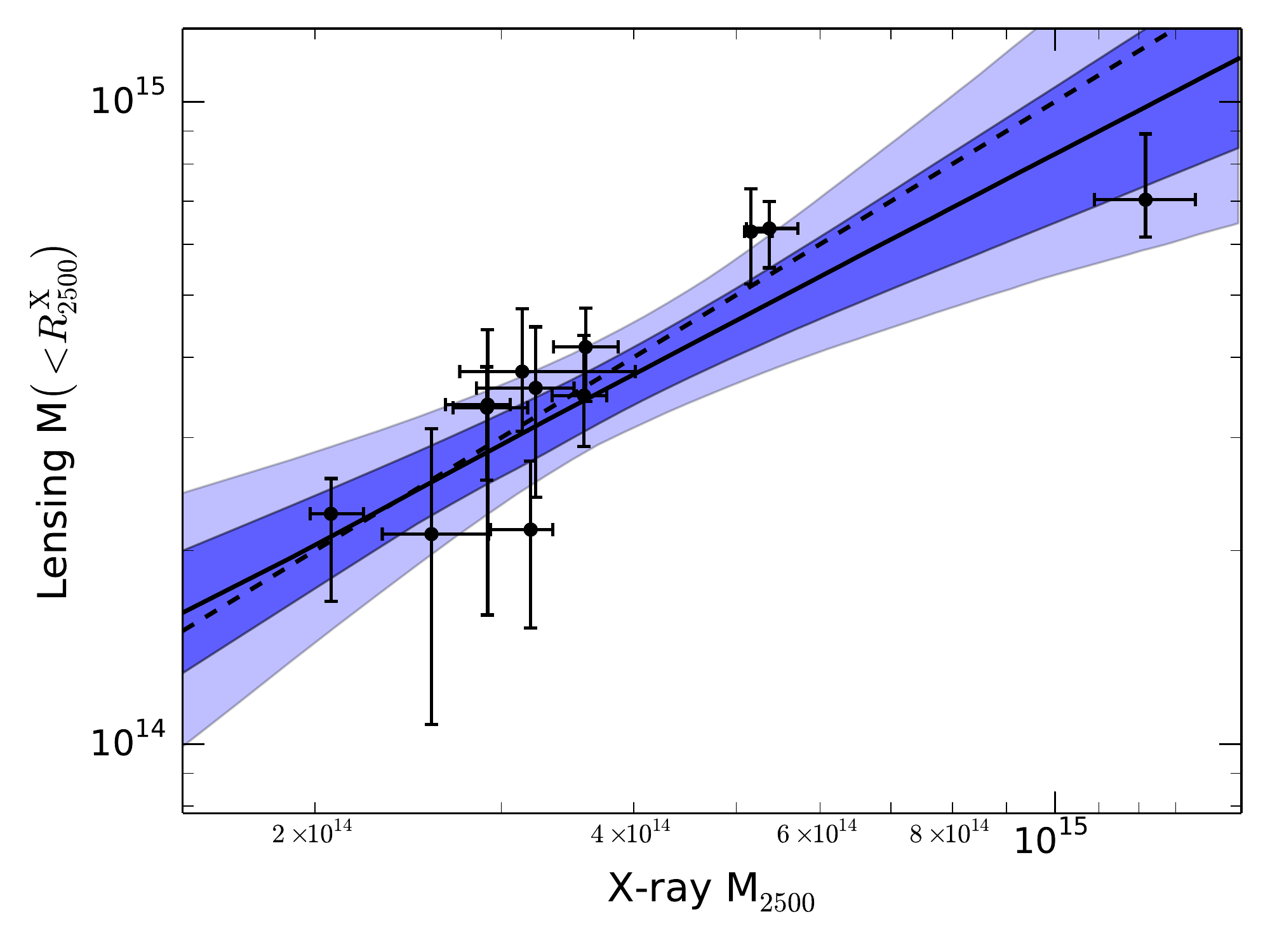}\\
\caption{Lensing mass versus X-ray mass. The dark and light blue bands are the 68\% and 95\% confidence intervals allowed by the power-law model fit to the data. The dashed black line is the one-to-one line. The magnitude of the intrinsic scatter has been marginalized over.}
\label{fig:powerlaw}
\end{figure}

We first examine if the lensing to X-ray mass ratio exhibits a mass dependence.
To do this, we extended the ratio model from Section~\ref{sec:xlratio} to include a power-law scaling between lensing and X-ray masses (again including intrinsic scatter), and model the intrinsic distribution of X-ray masses as a sum of Gaussians (see Section~\ref{sec:statmethods}).
We used the mean X-ray mass as the pivot point for the power-law.
A power law index of $\beta=1$ indicates no mass dependence and reduces to the original ratio model.
Figure~\ref{fig:powerlaw} shows the inferred power-law.
The statistical 68\% confidence interval for the power law index, with all other variables marginalized over, is $\beta = 0.78_{-0.14}^{+0.17}$.
The power law index $\beta$ is less than unity at less than $2\sigma$ significance.
We therefore find no statistical evidence for a trend with mass.
\footnote{It is an interesting to note that when we consider the ``relaxed WtG'' plus the marginal cluster samples, the power law index $\beta$ is measured to be $\beta = 0.69 \pm 0.13$, which is less than unity at more than  $2\sigma$ significance.
However, it is hard to interpret this measurement without a more robustly defined sample.}

We next check for an explicit redshift dependence in the lensing to X-ray ratio.
We extended the ratio model from Section~\ref{sec:xlratio} to include a term linear in cluster redshift, such that $\ln M_{2500}^L/M_{2500}^X = \alpha + \beta z$, using the mean redshift as the pivot point in the fit. 
We measured a best fit slope of $-0.80_{-0.69}^{+0.76}$, \ie a decreasing mass ratio towards higher redshifts.
However, the results are statistically consistent with a constant value at approximately $1\sigma$.



\section{Average Concentration of Massive, Relaxed Clusters}
\label{sec:ave_concen}

The mass--concentration relation for massive halos influences the measured lensing to X-ray mass ratio (as discussed in Section~\ref{sec:concentration}).
It also reflects the formation history of halos and carries some weak information on cosmology \citep{wechsler2002, ludlow2013}.
By design, the WtG mass measurements are minimally sensitive to assumptions about the cluster concentration (Section~\ref{sec:concentration})
However, some partial sensitivity is still present, which we use here to measure the average concentration of the massive, relaxed clusters in our sample.

As opposed to the rest of this study, for this exercise we jointly fit the concentration of six clusters with \pz measurements.
We limit ourselves to the \pz clusters to avoid systematic uncertainties from the ``contamination correction'' required in the color cut analysis \citep[see][]{paper3}.
The contamination correction alters the slope of the shear profile, and is therefore degenerate with the measured concentration.
We also remove the prior on concentration from Section~\ref{sec:lensing_masses}.

Since we designed our measurements to be insensitive to concentration, we do not expect to measure the concentration of individual clusters with any useful precision. 
In addition, we expect additional noise from triaxiality and large-scale structure \citep{bahe12}.
We therefore simultaneously model the population of clusters to measure an average concentration.
We assume that the concentrations of clusters will have a log-normal scatter around the average concentration. 
Specifically, our model consists of a mass ($M(<1.5\mathrm{Mpc})$) and concentration for each cluster, the mean of the log-normal distribution of concentrations, $\mu_c$, and the scatter in concentration measurements, $\sigma_c$.
We assume flat priors on $\mu_c$ and $\sigma_c$,  and restrict $\sigma_c$ to the range [0.286, 0.318], based on the largest mass bin reported in the simulation results of \citet{bahe12}.

We measure the average concentration to be $\mu_{c} = 3.0_{-1.8}^{+4.4}$.
While hardly discriminatory, this fit is consistent with our adopted prior on the concentration ($\mu_{c} = 4.6$) for measuring individual masses of clusters in previous sections.

Other groups have shown that WL observations can in principle achieve up to 10\% precision on average concentration measurements for an ensemble of clusters \citep{okabe13}.
However, this precision requires fitting NFW halo models to small radii (typically $\sim 150$kpc).
At these radii, the measured shear from these massive clusters exceeds $g\sim0.1$, and often exceeds $g\sim0.3$ \citep[see example shear profiles in][]{paper1}, while shear measurement codes have only been rigorously calibrated to shears of $g\sim0.05$ \citep{step2, great08, great10}.
Extending the shear calibration to this regime is an ongoing effort \citep{desc_whitepaper}.


\section{Discussion}
\label{sec:discussion}

\subsection{Interpreting the weak lensing to Chandra X-ray mass ratio}

We have measured a WL to \chandra{} X-ray mass ratio of \ratiomeas, with an additional 9\% systematic uncertainty from the lensing analysis. 
Combining the statistical and systematic uncertainties in quadrature gives $\approx12\%$ precision, making the measured mass ratio consistent with unity at $< 1\sigma$.

Hydrodynamic simulations have examined X-ray hydrostatic masses with different methodologies and baryonic physics prescriptions.
Simulations agree that, even for the most relaxed clusters, hydrostatic mass estimates should slightly under-report the true mass within $r_{2500}$, with $\lesssim 10\%$ bias  \citep{nagai07, lau09, rasia12, battaglia13}.
Therefore, these simulations would predict a lensing to X-ray mass ratio in the range of 1.0 - 1.1 at these radii.
Our results provide a 95\% confidence upper limit on the lensing to \chandra{} X-ray mass ratio of $\approx 1.15$.
This disfavors models with stronger violations of hydrostatic equilibrium in massive, relaxed clusters than predicted by current simulations.

However, we can only establish a truly robust limit on departures from HSE if we also have a firm understanding of the systematic uncertainties  in \chandra{} cluster mass measurements.
In particular, the temperature calibration for in-orbit X-ray detectors is an area of active research by the community (see \url{http://web.mit.edu/iachec/} and the documents linked therein).
A pessimistic bound on \chandra{} systematic errors would add an additional $\approx15\%$ systematic uncertainty to our results, permitting somewhat larger departures from HSE to be present in our clusters.
That would still disfavor a hydrostatic bias of $\approx20\%$ or more, which is one interpretation of the results from \citet{planck_clusters}.

Our results also allow the possibility that the current \chandra{} calibration may bias temperatures measured from continuum emission high, though we cannot place a strong constraint on the size of such a bias.
It is well documented that a discrepancy exists between the XMM-Newton and \chandra{} temperature measurements, with \chandra{} temperatures currently being higher than XMM \citep{ndg10, mahdavi13, schellenberger14}.
One examination of broad-band versus line-emission measurements of cluster temperatures suggests that the true temperatures may lie between those inferred from fitting the continuum emission measured by the two telescopes \citep{schellenberger12}.
If \chandra{} temperatures are indeed biased high, then so would be \chandra{} hydrostatic masses.
If the calibration bias exceeds the bias due to non-thermal support, this would result in a lensing to X-ray mass ratio of less than 1, consistent with our results.

\subsection{Comparison to Literature}

Other groups have previously compared lensing masses to X-ray hydrostatic masses measured with \chandra. 
In particular, \citet{donahue14} have  eleven clusters in common with our extended sample, and most of the raw \chandra{} and lensing data are the same.
\citet{donahue14} independently processed the \chandra{} data and adopted lensing masses from \citet{umetsu_clash}. 
They use a weighted mean to measure a lensing to X-ray ratio of $r=1.14\pm 0.09$ at 0.5Mpc,  approximately $r_{2500}$ for their clusters.

It is unclear why the results from \citet{donahue14} differ significantly from our measurements.
In part, the clusters analyzed in \citet{donahue14} are less relaxed on average than the clusters included in our sample, according to the measurements of Paper I.
However, even our measurements of the ``relaxed WtG plus marginal sample'' are still in some tension with \citet{donahue14}, given the nearly identical input data.
The differences are not obviously caused by discrepant lensing analyses.
 \citet{umetsu_clash} compared their lensing masses with Weighing the Giants lensing masses at 1.5Mpc and found general agreement between the two works, with a median ratio (WtG/CLASH) of 1.02 and a geometric mean of 1.10.
Note that taking the geometric mean as the actual WtG/Clash offset only exacerbates the discrepancy between \citet{donahue14} and this work.
\citet{umetsu_clash} measure an average concentration consistent with our assumed mass--concentration model, making a radius-dependent offset between WtG and CLASH unlikely.

Turning to the X-ray measurements, we find some discrepancies between the \citet{donahue14} analysis and our analysis.
For the eleven clusters in common, X-ray masses at $r_{2500}$  in that work are $\approx20\%$ lower than our measurements, with $\approx 15\%$ scatter.
We do not expect the \chandra{} calibration to be an issue, as \citet{donahue14} calibrate their data with CALDB 4.5.9, which should be comparable to the CALDB version 4.6.1 used in our work.
One clear methodological difference is that \citet{donahue14} bin their spectra and use a $\chi^2$ fitting statistic in their analysis, which is known to bias measurements of temperature when applied to relatively noisy data \citep{cash1979}. 
In contrast, our procedure minimally bins the spectra, and instead uses the C-statistic to correctly describe the Poisson nature of the observed counts.
We estimate that perhaps as much as half of the discrepancy between our X-ray masses and those of \citet{donahue14} could originate in this difference, based on tests fitting mock spectra with realistic backgrounds and signal-to-noise.
However, a full resolution of the discrepancy would require more detailed comparisons.

The works of \citet{vbe09} and \citet{israel14} also feature WL calibrations of \chandra{} derived X-ray masses.
However, no clusters are in common between this work and \citet{israel14}, while only 2 clusters are in common with the \citet{vbe09} calibration sample (Benson, priv comm).
Furthermore, both works use a custom \chandra{} temperature calibration from \citet{vikhlinin_calib} that is not publicly available, and neither work restricts itself to relaxed clusters.

Both \citet{mahdavi13} and \citet{zhang10} compare WL masses to XMM-Newton derived hydrostatic masses for samples of relaxed clusters. 
However, few clusters are in common between those samples and this work.
Also, both of those works use lensing data, from \citet{hoekstra12} and \citet{okabe_masses}, respectively, that are known to sharply disagree with Weighing the Giants masses.
Both analyses have been revised in the interim \citep[see ][]{hoekstra2015, okabe13}.
A comparison of the mass ratio results would therefore be of limited value.

Insufficient information is currently available to generalize measurements of the lensing to X-ray mass ratio from one X-ray telescope or X-ray calibration version to others.
Ideally, the community needs a large sample of relaxed clusters that are consistently analyzed with each calibration version and both \chandra{} and XMM.
This would facilitate comparisons among X-ray based results measured at different times.


\subsection{Cosmology Dependence}
\label{sec:cosmo_dependence}

\begin{figure}
\includegraphics[width=\columnwidth]{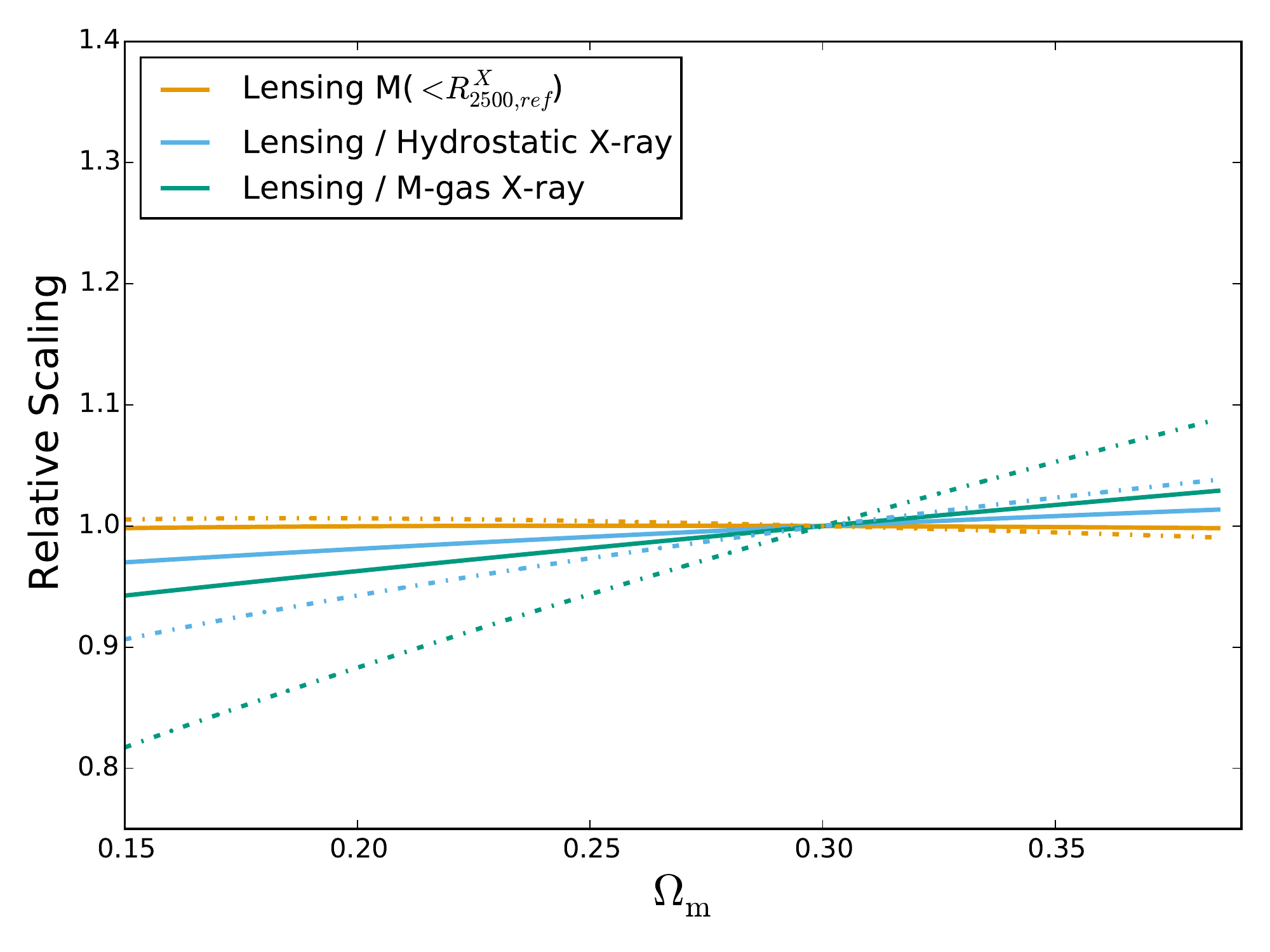}\\
\includegraphics[width=\columnwidth]{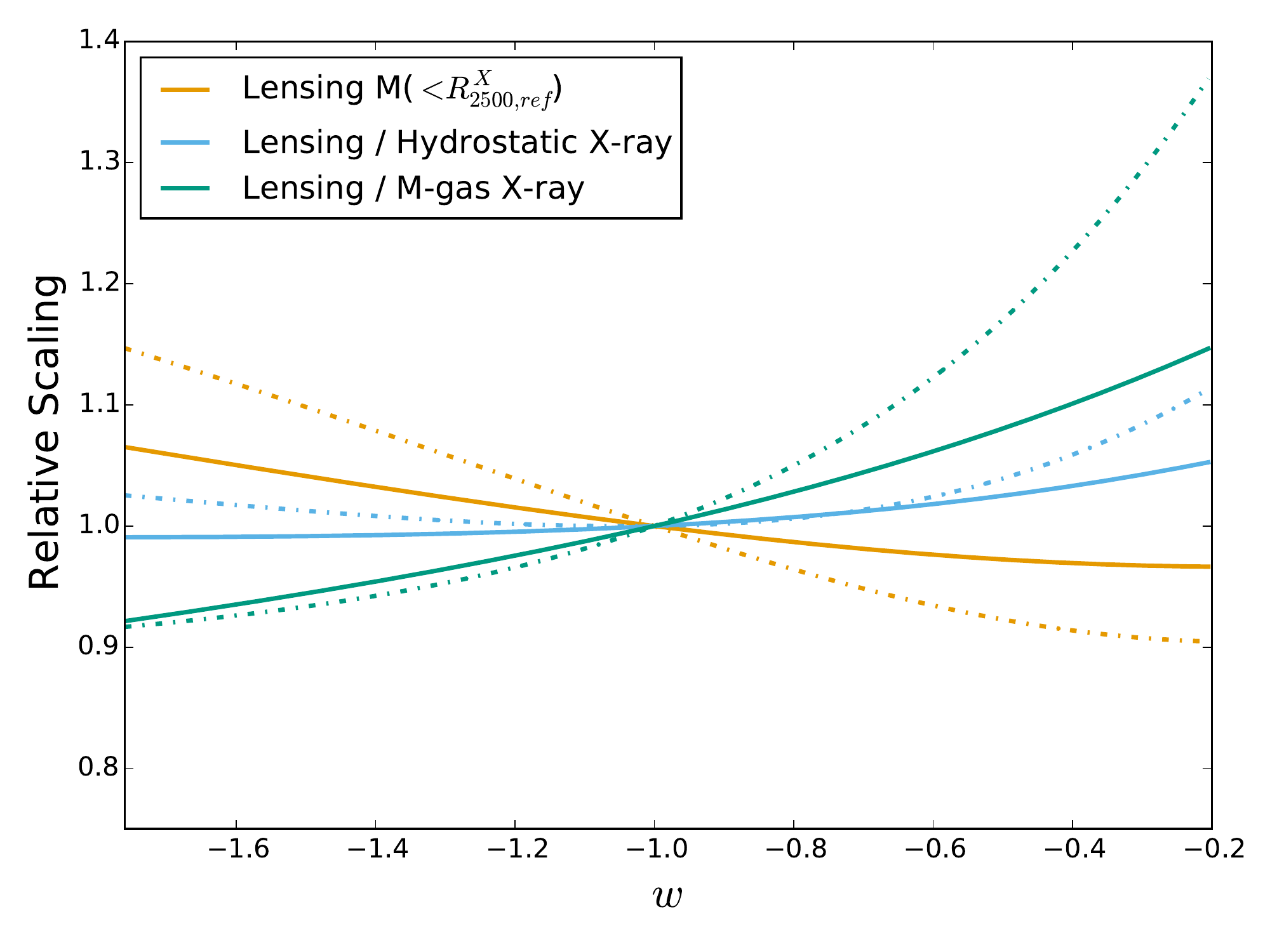}
\caption{Plotted curves show how lensing masses in fixed apertures (yellow) and the WL to X-ray mass ratio (hydrostatic masses in blue and gas mass proxies in green) scale with $\Omega_m$ (top) and $w$ (bottom) as a function of X-ray mass type (colors) and by cluster redshift (solid: z=0.25, dashed: z=1.0), as given by equation~\ref{eq:cosmo_scaling}.  All curves are normalized to $\Omega_m = 0.3$, $w=-1$. Whereas the lensing to X-ray hydrostatic mass ratio does not vary with $\Omega_m$ at a level significant for this study, future work calibrating X-ray mass measurements for high redshift clusters will need to account for this dependence.}
\label{fig:cosmo_dependence}
\end{figure}

Both the X-ray and lensing masses in this analysis depend on cosmology.
The X-ray hydrostatic mass depends on the physical volume filled by gas of a given temperature, resulting in a dependence on the angular diameter distance $D$ to the cluster.
Lensing masses, on the other hand, depend on both the angular diameter distance to the cluster and on the angular diameter distances from the observer and cluster to the background lensed galaxy population.
For example, the 3D lensing mass within a fixed angular aperture will scale approximately as 
\begin{equation}
M_L(<R) \propto D_{\mathrm{lens}}^2\Sigma_{\mathrm{crit}} = \frac{D_{\mathrm{lens}}D_{\mathrm{source}}}{D_{\mathrm{lens-source}}}.
\label{eq:cosmo_scaling}
\end{equation}
Equation~\ref{eq:cosmo_scaling} is exact for a Single Isothermal Sphere (SIS) model with all sources at one redshift.
The actual scaling depends on the type of mass calculated (SIS vs.\ NFW, aperture statistic vs.\ profile fit), the angular size of the measurement aperture, the functional form used to model the mass distribution, and the redshift distribution of lensed galaxies.
For flat models, this scaling is roughly independent of $\Omega_m$, but can vary by up to 10\% with $w$, as seen in figure~\ref{fig:cosmo_dependence}.
Since lensing masses depend on cosmology differently from X-ray hydrostatic masses, the lensing to X-ray mass ratio will also have a cosmology dependence.
To check this toy model, we re-evaluated the lensing to X-ray mass ratio from section~\ref{sec:xlratio} at four cosmologies, assuming flatness, that encompass the $2\sigma$ allowed region of the \fgas experiment from Paper II, $\Omega_m = [0.21, 0.40]$ and $w = [-2.0, -0.35]$.
The mass ratio shifted as expected for clusters at $z \lesssim 0.5$, changing by less than $\approx 5\%$.

X-ray hydrostatic masses are only one example where a calibration with lensing is cosmology dependent. 
X-ray gas mass and $Y_x$ mass proxies measured within a fixed aperture scale as $D^{5/2}$.
Therefore the lensing calibration of these mass proxies will be even more sensitive to cosmology than hydrostatic masses.
While this sensitivity to cosmological parameters is perhaps negligible for today's experiments, next generation experiments and measurements of high redshift clusters will need to account for it.
Our solution is to explicitly include the lensing data in the cosmological likelihood, by effectively fitting for lensing masses simultaneously with cosmology (see Section~\ref{sec:statmethods}). 
In this way, we account for the exact cosmology dependence of the lensing masses and the relevant X-ray mass proxy.

\begin{figure}
\includegraphics[width=\columnwidth]{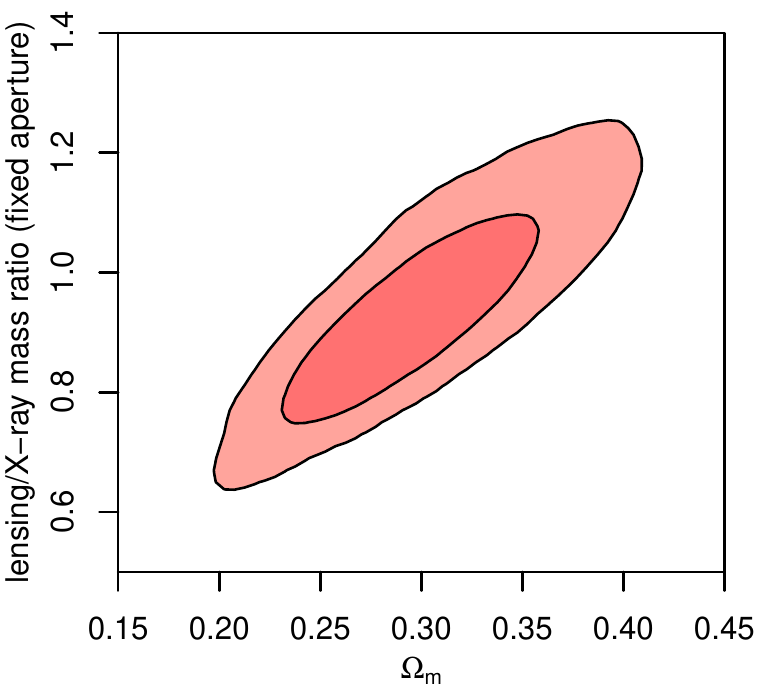}
\caption{
  Joint $1\sigma$ and $2\sigma$ confidence regions for $\Omega_\mathrm{m}$ and the lensing to X-ray mass ratio for the \fgas{} measurement, marginalized over the full model used in Paper~II (specifically for a flat, constant-$w$ cosmology). In the \fgas model, the mass ratio is proportional to $\Omega_\mathrm{m}D^{3/2}$, whereas the extent of the allowed region is primarily set by the precision of the mass ratio measurement. As these results come directly from the analysis of Paper~II, they use the May 2012 \chandra{} calibration (CALDB 4.4.10) and do not include the correction for the lensing mass model. This has a small effect on the mass ratio; see Sections~\ref{sec:lensing_masses} and \ref{sec:calib_dependence}.}
\label{fig:cosmo2}
\end{figure}

As an example of why one should explicitly include the lensing data in a cosmological likelihood, we investigate how the cosmology dependence of the lensing to X-ray ratio manifests itself in the \fgas test.
In the context of the \fgas test, the mass ratio scales as $K(z) \propto \Omega_\mathrm{m}D^{3/2}$ (see equation 2 from Paper II, where $K(z)$ is the mass ratio).
Figure~\ref{fig:cosmo2} shows the degeneracy between the mass ratio and $\Omega_\mathrm{m}$ that appears in practice when fitting the full model of Paper~II (for a flat constant-$w$ model), reflecting the $\Omega_\mathrm{m}D^{3/2}$ dependence.
Neglecting the cosmology dependence of the lensing calibration would introduce artificial priors on cosmological constraints from the \fgas test (or any other cosmological test that uses lensing as a calibrator), potentially biasing results and underestimating uncertainties.
Conversely, the cosmological model space explored has an effect on the inferred mass ratio, albeit negligible with the current statistical uncertainties.
We find that measured lensing to X-ray mass ratio shifts by $\Delta r = 0.01$ between a flat constant-$w$ model and a $\Lambda$CDM model with curvature.

\subsection{Blind Analysis}
\label{sec:blindanalysis}

Our aim in this work is to calibrate X-ray hydrostatic mass measurements for relaxed clusters to $\sim10\%$ accuracy, a level  where ``observer's bias'' must be guarded against.
There is a clear expectation in the literature that X-ray hydrostatic masses should be biased low compared to the true mass \citep{nagai07}.
One group previously re-calibrated \chandra{} X-ray masses downwards by 15\% to match these expectations \citep{mahdavi13}.

To combat observer's bias, our lensing and X-ray analyses were completed independently and by different members of the team.
For the X-ray analysis, gas and total mass profiles for individual clusters were blinded with random offsets until after the relaxed cluster sample was identified and their profiles analyzed (Paper~II). 
For the lensing analysis, fit procedures and systematic error estimates were fixed before comparisons to literature lensing values occurred \citep{paper3}.
Comparison between lensing and X-ray hydrostatic masses occurred only after the sample of relaxed clusters was finalized.

For full disclosure, we briefly discuss changes that occurred in the analysis during and after the unblinding process.
Following the selection of relaxed clusters, but before \fgas values were unblinded, we computed the lensing to X-ray mass ratio for 6 clusters using the \pz likelihood from \citet{paper3}, measuring a ratio of $r=0.95\pm 0.14$.
Following the unblinding of \fgas values, we received the referee report for \citet{paper3}; however the report did not result in changes to the analysis.
Upgrades to the color-cut analysis described in Section~\ref{sec:data}, most notably marginalizing over a prior on concentration appropriate for relaxed clusters, were then finalized before lensing and X-ray masses for the full sample of 12 clusters were compared, resulting in a ratio of $r = 0.959^{+0.06}_{-0.06}$.
Subsequently, we concluded that the cosmology dependence of the lensing to X-ray mass ratio must be included in the \fgas analysis, resulting in the lensing likelihood for color-cut masses being incorporated into the \fgas analysis (see Section~\ref{sec:cosmo_dependence}).
The \chandra{} effective area calibration was also updated, to CIAO version 4.6.1 and CALDB version 4.6.2 (Section~\ref{sec:calib_dependence}).
During that process, we fixed a bug in the importance sampling algorithm used to evaluate the ratio at fixed cosmology (Section~\ref{sec:xlratio}).
Immediately before submission, new simulations results became available indicating that the lensing masses were underreported by a few percent (Section~\ref{sec:lensing_masses}), after which a correction factor was introduced.
The cumulative effect of these changes shifted the ratio for our reference cosmology to \ratiomeas.


\section{Summary}
\label{sec:summary}

In this work, we have calibrated \chandra{} hydrostatic masses for relaxed clusters with accurate weak lensing measurements from the Weighing the Giants project \citep{paper1, paper2, paper3}.
Specifically, we measured the combined astrophysical and instrumental bias present for the \citet{mantz_fgas} sample.
Our approach fully captures lensing and X-ray measurement uncertainties, correlations between the measurements when using a common analysis aperture, the intrinsic scatter between X-ray and lensing masses, and the cosmology dependence of both.
For a fixed cosmology and a sample of 12 clusters, we measure a lensing to X-ray mass ratio of $r=0.96$ within the \chandra-determined $r_{2500}$ aperature with a statistical precision of 9\% and additional systematic uncertainty from lensing of 9\%, for a combined precision of $\approx12\%$.
These results are for the specific \chandra{} calibrations tested, and cannot be easily extrapolated to other versions of the \chandra{} calibration.
We find that our results are robust to perturbations in the sample selection and lensing analysis.

We interpret the measured X-ray to lensing ratio as ruling out large departures from hydrostatic equilibrium, on the order of tens of percent, at the measurement radii in our relaxed target clusters. 
Since the effects of biases in the X-ray effective area calibration are degenerate with biases due to non-thermal pressure support, more robust constraints on the level of non-thermal support from lensing and X-ray data will not be possible without an independent constraint on the X-ray calibration, \eg from precisely measured emission line ratios.

By measuring the lensing to X-ray mass ratio within the cosmological framework of Paper II, we are able to render the results of that work independent of overall shifts in the \chandra{} temperature calibration.
We are also able to naturally capture the cosmology dependence of the lensing to X-ray mass ratio by simultaneously modeling the lensing signal for the 12 clusters in this sample with the rest of the \fgas data set.
The improvements to cosmological constraints that the lensing enables are significant, as discussed in detail in Paper II.

There are clear opportunities for improvement of the lensing-based calibration for the \fgas experiment (as well as cluster counts; see discussion in \citealt{paper4}) in the next few years.
Efforts are under way to improve key lensing systematics, most notably with measuring shear \citep{great10} and with mass modeling of clusters \citep{desc_whitepaper}.
With these improvements coming into place, the remaining limitation to an improved measurement of the lensing to X-ray mass ratio is the number of clusters with sufficient optical filter coverage for \pz measurements, which eliminate a key systematic in the lensing masses used in this work \citep[see][]{paper3}.
Expanding the sample of \pz cluster masses from the current 6 clusters to $\sim24$ clusters would already improve the mass ratio precision, and thus constraints on $\Omega_m$, by $\approx 30\%$.


\section*{Acknowledgments}

We thank our other collaborators in the Weighing the Giants series, including Pat Burchat, Mark Allen, and David Donovan.
DEA would like to thank Lorenzo Lovisari and Tim Schrabback for helpful discussions.
SH thanks MPA and RZG for hospitality and for providing access to their computing facilities and the MXXL simulation data.

DEA recognizes the support of the German Federal Ministry of Economics and Technology (BMWi) under project 50 OR 1210.
This work is supported in part by the U.S. Department of Energy under
contract number DE-AC02-76SF00515. This work was also supported by the
National Science Foundation under Grant No. AST-0807458. AM acknowledges the
support of NSF grant AST-1140019.  The authors
acknowledge support from programs HST-AR-12654.01-A,
HST-GO-12009.02-A, and HST-GO-11100.02-A provided by NASA through a
grant from the Space Telescope Science Institute, which is operated by
the Association of Universities for Research in Astronomy, Inc., under
NASA contract NAS 5-26555. This work is also supported by the National
Aeronautics and Space Administration through Chandra Award Numbers
TM1-12010X, GO0-11149X, GO9-0141X , and GO8-9119X issued by the
Chandra X-ray Observatory Center, which is operated by the Smithsonian
Astrophysical Observatory for and on behalf of the National
Aeronautics Space Administration under contract NAS8-03060.  The Dark
Cosmology Centre (DARK) is funded by the Danish National Research
Foundation.

Based in part on data collected at Subaru Telescope (University of
Tokyo) and obtained from the SMOKA, which is operated by the Astronomy
Data Center, National Astronomical Observatory of Japan.  Based on
observations obtained with MegaPrime/MegaCam, a joint project of CFHT
and CEA/DAPNIA, at the Canada-France-Hawaii Telescope (CFHT) which is
operated by the National Research Council (NRC) of Canada, the
Institute National des Sciences de l'Univers of the Centre National de
la Recherche Scientifique of France, and the University of Hawaii.
This research used the facilities of the Canadian Astronomy Data
Centre operated by the National Research Council of Canada with the
support of the Canadian Space Agency.  This research has made use of
the VizieR catalogue access tool, CDS, Strasbourg, France. Funding for
SDSS-III has been provided by the Alfred P. Sloan Foundation, the
Participating Institutions, the National Science Foundation, and the
U.S. Department of Energy Office of Science. The SDSS-III web site is
http://www.sdss3.org/. This research has made use of the NASA/IPAC
Extragalactic Database (NED), which is operated by the Jet Propulsion
Laboratory, Caltech, under contract with NASA.

\bibliography{refs.bib}

\appendix

\section{Ratio Model with Intrinsic Scatter and Common Apertures}
\label{sec:likelihood}

There are multiple, mathematically equivalent ways to implement a ratio model that includes intrinsic scatter and accounts for correlations induced from a shared measurement aperture. 
For completeness, we describe here one of two ways that was used in this work. 

We need to evaluate the posterior probability function $P(r,\sigma_{\mathrm{int}}| \mathcal{D}_X,\mathcal{D}_L)$, where $r$ is the lensing to X-ray mass ratio, $\sigma_{\mathrm{int}}$ is the magnitude of intrinsic scatter present in the sample, $\mathcal{D}_X$ is the X-ray data vector, and $\mathcal{D}_L$ is the lensing data vector. 
For each cluster, we first constrain the parameters of an NFW mass model with the X-ray and lensing data independently, thereby obtaining samples from the probability distributions for the mass $M_{i}$ and  concentration $C_{i}$ from each data set, $P(M_{i,X},C_{i,X}|\mathcal{D}_{i,X})$ and $P(M_{i,L},C_{i,L}|\mathcal{D}_{i,L})$. 
Note that we also include a prior on the concentration measured by lensing, $P(C_{i,L})$, which we will ignore in the following description.
The X-ray or lensing $r_{2500}$ mass within the measurement aperture is a deterministic function of the NFW halo parameters: $M_{i,L}^\mathrm{AP} = f(M_{i,L},C_{i,L}, M_{i,X},C_{i,X})$ and $M_{i,X}^\mathrm{AP} = g(M_{i,L},C_{i,L}, M_{i,X},C_{i,X})$, respectively. 
The exact parameter dependence of each depends on the chosen measurement aperture.
This accounts for the use of a common measurement aperture.

The posterior probability function $P(r,\sigma_{\mathrm{int}}| \mathcal{D}_X,\mathcal{D}_L)$ can then be expanded as
\begin{multline}
P(r, \sigma_{\mathrm{int}}|\mathcal{D})  \propto \prod_i\int \dif M_{i,X}\dif C_{i,X}\dif M_{i,L}\dif C_{i,L} \, P(r, \sigma_{\rm int} | M_{i,X}^\mathrm{AP}, M_{i,L}^\mathrm{AP})\\
P(M_{i,X},C_{i,X}|\mathcal{D}_{i,X})P(M_{i,L},C_{i,L}|\mathcal{D}_{i,L}),
\label{eq:posterior}
\end{multline}
where the deterministic calculations of $M_{i,X}^\mathrm{AP}$ and $M_{i,L}^\mathrm{AP}$ have been suppressed. In our model with log normal scatter,
\begin{equation}
P(r, \sigma_{\rm int} | M_{i,X}^\mathrm{AP}, M_{i,L}^\mathrm{AP}) = \frac{1}{r\,\sigma_{\rm int}\sqrt{2\pi}}e^{-\frac{\left(\ln rM_{i,X}^\mathrm{AP} - \ln M_{i,L}^\mathrm{AP}\right)^2}{2\sigma_{\rm int}^2}}.
\end{equation}

We evaluate the integrals in equation~\ref{eq:posterior} numerically using Monte Carlo integration, transforming the integrals into weighted averages over the pre-computed samples from $P(M_{i,X},C_{i,X}|\mathcal{D}_{i,X})$ and $P(M_{i,L},C_{i,L}|\mathcal{D}_{i,L})$.


\label{lastpage}
\end{document}